Revision 2

# Evidence of Permian magmatism in the Alpi Apuane Metamorphic Complex (Northern Apennines, Italy): new hints for the geological evolution of the basement of the Adria plate

Simone Vezzoni[1], Cristian Biagioni[1,*], Massimo D'Orazio[1], Diego Pieruccioni[1], Yuri Galanti[1], Maurizio Petrelli[2,3], and Giancarlo Molli[1]

[1] *Dipartimento di Scienze della Terra, Università di Pisa, Via S. Maria 53, I-56126, Pisa, Italy*

[2] *Dipartimento di Fisica e Geologia, Università di Perugia, Piazza dell'Università, I-06123 Perugia, Italy*

[3] *Istituto Nazionale di Fisica Nucleare, Sezione di Perugia, Via Alessandro Pascoli 23c, I-06123 Perugia, Italy*

*Corresponding author: cristian.biagioni@unipi.it

## 1. Introduction

The Paleozoic basement of the Northern Apennines (central-northern Italy) has been described from few small outcrops, with the only exception being represented by the Alpi Apuane where large outcrops (ca. 60 km$^2$) of pre-Triassic formations are widely exposed. During the Alpine tectonic evolution, these formations were part of the basement of the Adria plate which, according to different authors (e.g., Stampfli and Borel, 2002; Stampfli et al., 2002; von Raumer et al., 2013), was located, in the Paleozoic paleotectonic frame, along the northern margin of Gondwana, therefore recording the tectono-sedimentary and magmatic processes associated with its geodynamic evolution (e.g., Sirevaag et al., 2017). Indeed, during the late Neoproterozoic-middle Paleozoic, the northern margin of Gondwana was modified first by crustal accretion and then by repeated rifting and associated intense magmatism (e.g., Romeo et al., 2006; von Raumer et al., 2013). In the late Paleozoic, many Gondwana-derived continental fragments, among which part of the future Adria plate, were incorporated into the Variscan orogeny, deformed at different *P-T* conditions, and affected by extensive magmatism occurred during both the collisional and post-collisional phases (e.g., Kroner and Romer, 2013). Such events have been differently recorded in the Paleozoic basement units of several circum-Mediterranean chains (von Raumer et al., 2013).

Several difficulties in the interpretation of this record in the Northern Apennine basement have been encountered by previous authors, mainly owing to the small and scattered outcrops, the rarity of fossiliferous layers in metasedimentary units, and the Alpine metamorphic overprint, usually obliterating or masking the older structures. Moreover, very few geochronological data are available to date. Indeed, the present knowledge of the basement stratigraphy derives mainly from geological, petrographical, and geochemical data, coupled with a correlation with the better known and well-dated Paleozoic sequences of central Sardinia, allowing some tentative stratigraphic reconstructions (e.g., Pandeli et al., 1994 and references therein). Only recently, Musumeci et al. (2011), Sirevaag et al. (2017), and Paoli et al. (2017) reported U-Pb dating on detrital as well as magmatic zircon from rocks belonging to the basement of the Northern Apennines, giving a significant contribution to its stratigraphy.

The stratigraphic succession of the Alpi Apuane suffers the same lack of geochronological data common to the other outcrops of Northern Apennines. In addition to the biostratigraphical dating of Vai (1972) and Bagnoli and Tongiorgi (1980), the only available geochronological data were given by Paoli et al. (2017). These authors suggested early Cambrian depositional ages for both the "*Filladi inferiori*" (Lower Phyllites) and "*Filladi superiori*" (Upper Phyllites) Fms, and a crystallization age of 457 ± 3 Ma for the "*Porfiroidi e scisti porfirici*" (Porphyroids and Porphyritic Schists) Fm. The Alpi Apuane represent not only a privileged area for the study of the older formations recording the evolution of the northern Gondwana margin up to their incorporation into the Variscan orogenic belt, but also for the investigation of several polymetallic ore deposits, mostly related to hydrothermal circulation within the Paleozoic basement.

Within the frame of an ongoing study of the mineralogy, geochemistry, and structural setting of such ore deposits from the southern Alpi Apuane, a close spatial relationship between the pyrite ± baryte ± iron oxide ± (Pb-Zn-Ag) ore deposits (D'Orazio et al., 2017) and felsic, tourmaline-rich, meta-igneous rock bodies belonging to the Paleozoic basement was observed. Such bodies may achieve a length of more than 1 km and an apparent thickness larger than 100 m (Pieruccioni et al., 2018). This peculiar association, as well as the massive nature of these rock bodies, completely different from the surrounding schists, promoted a field survey of these rocks, known since the end of the 19$^{th}$ Century (e.g., Lotti, 1882), integrated by petrographic, geochemical, and geochronological data. In this paper, the results of this study are reported, emphasizing both the geodynamic significance of such rocks in the framework of the Adria plate basement evolution and their implications for the hydrothermal processes affecting the Paleozoic sequences of the Alpi Apuane.

## 2. Geological framework of Alpi Apuane

*2.1. Geological background*

The Northern Apennines are an orogenic belt formed during the Oligocene-Miocene collision between the Corsica-Sardinia microplate and the Adria plate. As a consequence of this collision, stacked slices of the former Adria plate continental margin (Tuscan and external foreland tectonic units) lie below the ocean-derived, ophiolite-bearing, Ligurian units. The Tuscan units locally preserve the Paleozoic basement overlain by Mesozoic and Cenozoic cover sequences (Molli, 2008 and references therein).

In the Alpi Apuane area, a large tectonic window exposes the lowermost units of the inner side of the Northern Apennines, represented by the Apuane and Massa units (Fig. 1). These units were affected by regional metamorphism under moderate pressure greenschist facies conditions during late Oligocene-early Miocene (e.g., Di Pisa et al., 1985; Franceschelli et al., 1986; Jolivet et al., 1998; Franceschelli et al., 2004; Fellin et al., 2007, and references therein). The present structural setting is the result of two main tectono-metamorphic events (D1 and D2, according to Carmignani and Kligfield, 1990), related to the deformation of the Adria plate continental margin during the continent-continent collision and the syn- to post-contractional exhumation (e.g., Molli, 2008).

The studied area (Fig. 2) is located in the southern part of the Alpi Apuane and extends along a narrow SW-NE belt, ca. 10 km in length, running from the village of Valdicastello Carducci, near the small city of Pietrasanta, to the hamlet of Fornovolasco. Its position completely overlaps with the discontinuous mineralized belt known for the occurrence of pyrite ± baryte ± iron oxide ± (Pb-Zn-Ag) ore deposits (D'Orazio et al., 2017). In this area, the metamorphic rocks belonging to the Apuane Unit crop out below the Tuscan Nappe in two tectonic windows, i.e., the Sant'Anna and Fornovolasco tectonic windows, in the SW and NE corners of this belt, respectively, and in the Stazzemese area, in the central part of the studied area. The Apuane Unit is here characterized by km-scale recumbent D2 isoclinal folds and tectonic slices, with rocks of the Paleozoic basement locally overlying the younger Meso-Cenozoic metasedimentary sequences. The Paleozoic basement cropping out in the studied area is composed of metasiliciclastic rocks with local intercalations of lenses of porphyritic tourmaline-rich rocks and shows some peculiar features with respect to the other outcrops of basement rocks occurring in the Alpi Apuane. Indeed, it is characterized by a widespread tourmalinization and by the occurrence of several small ore deposits embedded in the basement or close to the contact with the Triassic metadolostone ("*Grezzoni*" Fm). As a consequence, some authors proposed different interpretations of this lithostratigraphic succession, in some cases using specific formational names to identify it (e.g., "Fornovolasco Schists" – Pandeli et al., 2004). Other authors proposed a correlation between this peculiar succession and the remaining part of the pre-Triassic basement of the Apuane Unit, identifying the metasedimentary rocks with the "*Filladi inferiori*" Fm and the interlayered lenses of porphyritic tourmaline-rich rocks with the "*Porfiroidi e scisti porfirici*" Fm (Carmignani et al., 1976; Pandeli et al., 2004).

*2.2. The metavolcanic rocks*

The occurrence of metavolcanic rocks within the Paleozoic basement of Alpi Apuane has been known since the end of the 19$^{th}$ Century, even if the first comprehensive petrographical study was performed by Bonatti (1938).

The "*Porfiroidi e scisti porfirici*" Fm (Barberi and Giglia, 1965) represents the main felsic metavolcanic rocks ("*Porfiroidi*") and their subaerial reworking products ("*Scisti Porfirici*") occurring in the basement of Alpi Apuane. These rocks are characterized by a primary porphyritic texture and well-developed foliations related to both the Variscan and Alpine tectono-metamorphic events (Conti et al., 1991). They show a distinct augen texture, with large magmatic phenocrysts of quartz (usually with magmatic embayment) and K-feldspar, in a fine-grained matrix formed by muscovite, quartz, and chlorite (Barberi and Giglia, 1965). Chemical data indicate a rhyolitic/dacitic composition for the volcanic protolith (e.g., Puxeddu et al., 1984).

The age of these rocks has long been debated. Indeed, different authors proposed ages ranging from Permian (e.g., Barberi and Giglia, 1965) to middle Ordovician (e.g., Gattiglio et al., 1989). This latter age was proposed correlating the "*Porfiroidi e scisti porfirici*" Fm with similar metavolcanic products occurring in Sardinia. Recently, Paoli et al. (2017) confirmed such an age attribution through U-Pb zircon dating, indicating a crystallization age of 457 ± 3 Ma, similar to that reported for the Ortano Porphyroids (Elba Island) by Musumeci et al. (2011) and Sirevaag et al. (2017) and for the calc-alkaline effusive magmatic rocks of the Variscan chain occurring in the Paleozoic basement of Sardinia, dated at 465 ± 1 Ma (Oggiano et al., 2010) and 448 ± 5 Ma (Cruciani et al., 2013).

Locally, the "*Porfiroidi e scisti porfirici*", as well as the "*Filladi inferiori*" Fms, host small bodies of greenish and grey-greenish metabasites, tentatively assigned to a late to post-Ordovician magmatism (e.g., Conti et al., 1988, 1993).

The lenticular bodies of porphyritic tourmaline-rich rocks occurring within the Paleozoic phyllites in the southern Alpi Apuane have attracted the attention of several geologists during the second half of the 19$^{th}$ Century and the first decades of the 20$^{th}$ Century, mainly owing to their massive nature, contrasting with the surrounding strongly foliated schists. The first petrographic description was given by Bonatti (1933), who described the small outcrop of Le Casette, near the hamlet of Fornovolasco (Fig. 2). This lithology is mainly composed of quartz (sometimes showing magmatic embayment), tourmaline, feldspar, and muscovite. Accessory minerals are represented by apatite and rutile. According to Bonatti (1933), this rock could be interpreted as a meta-tufite. Since then, the interpretation of the stratigraphic significance of these tourmaline-rich rocks has been debated. For instance, Conti et al. (2012) described such rocks as a potential Tertiary aplite, whereas Pandeli et al. (2004) related them to the middle Ordovician "*Porfiroidi*". Finally, following the new data obtained during the present study, Pieruccioni et al. (2018) proposed a distinction between these rock bodies and the "*Porfiroidi*", defining the "Fornovolasco Metarhyolite" Fm.

## 3. Samples and methods

The samples studied in this work represent all the major known occurrences of the rocks belonging to the Fornovolasco Metarhyolite Fm in the Alpi Apuane (Fig. 2; Table 1). For the sake of comparison, three samples of "*Porfiroidi*" (Fig. 1; Table 1), showing the same petrographic features previously described by Barberi and Giglia (1965), were geochemically characterized.

The petrographic features of the rock samples were investigated by optical microscopy integrated by Scanning Electron Microscopy (SEM) observations. A SEM Philips XL 30 operating at 20 kV accelerating voltage and 5 µm beam diameter coupled with an energy-dispersive X-ray fluorescence spectrometer EDAX PV 9900 at the Pisa University's Dipartimento di Scienze della Terra was used. Electron-microprobe analyses were performed to fully characterize their mineralogy. Chemical data have been collected using an automated JEOL 8200 Super Probe at the Milano University's Dipartimento di Scienze della Terra "Ardito Desio". Operating conditions were 15 kV accelerating voltage, beam current of 5 nA, and 3 μm beam size, using wavelength dispersive spectrometry (WDS). A φ(ρZ) routine was used for matrix correction (Pouchou and Pichoir, 1991). Standards (element, emission line) were: omphacite (Na$K\alpha$), K-feldspar (K$K\alpha$), rhodonite (Mn$K\alpha$), olivine (Mg$K\alpha$), grossular (Al$K\alpha$, Si$K\alpha$, Ca$K\alpha$), fayalite (Fe$K\alpha$), apatite (P$K\alpha$), ilmenite (Ti$K\alpha$), nickeline (Ni$K\alpha$), metal Cr (Cr$K\alpha$), hornblende (F$K\alpha$), celestine (Sr$L\alpha$), metal V (V$K\alpha$), and sanbornite (Ba$L\alpha$). Counting times were 30 s for peak and 10 s for left and right backgrounds.

Chemical analyses of whole rocks were performed at ACTLABS (Ancaster, Ontario, Canada). Major and trace elements were determined by ICP-OES and ICP-MS, respectively, following lithium metaborate/tetraborate fusion and dissolution with diluted $HNO_3$.

Geochronological data were obtained by zircon U-Pb analyses on four samples (FVb13, FVc1, POL1, and SAS1). After crushing and sieving, zircon crystals were concentrated from the 90-250 μm grain size interval using standard separation techniques. About 100 zircon crystals for each sample were embedded in epoxy resin, ground to expose approximately mid-section of crystals, and finally polished with 0.25 μm alumina paste. Crystals for geochronological analyses were selected using SEM-cathodoluminescence (SEM-CL) taking into account the occurrence of inclusions, cores and/or rims and compositional zoning. Laser Ablation-Inductively Coupled Plasma-Mass Spectrometry (LA-ICP-MS) U-Pb analyses were performed at the Perugia University's Dipartimento di Fisica e Geologia using an iCAPQ Thermo Fisher Scientific, quadrupole-based, ICP-MS coupled to a G2 Teledyne Photon Machine ArF (193 nm) LA system. Uranium-Pb zircon analyses were calibrated with the international reference material zircon 91500 using a spot size of 25 μm and the Plešovice zircon had been used as quality control (Sláma et al., 2008). Data reduction was performed by the VizualAge protocol (Petrus and Kamber, 2012). In particular, raw signal counts and their ratios were carefully monitored in order to exclude from age calculations portions that may be contaminated by inclusions and/or spurious peaks. In addition, complex signals that may represent multiple ages had been carefully inspected to avoid misleading interpretations of the profiles. Net background-corrected count rates for each isotope were used for calculation (for further details, Petrelli et al., 2016). The analytical data were treated using the Isoplot Excel toolkit (Ludwig, 2003). The $^{206}Pb/^{238}U$ ages were used for probability plots.

## 4. Results

### 4.1. Field study

The rocks belonging to the Fornovolasco Metarhyolite Fm (Figg. 3, 4) are granular to porphyritic and white to dark-grey in color. Locally, tourmaline spots, up to some cm in size and with a leucocratic rim halo, occur (Fig. 3a). This latter feature is particularly developed in the large outcrop of the Boscaccio locality, east of Fornovolasco, whereas it is rarer in the other outcrops. The most striking feature is the massive nature of these rocks (Fig. 3). Only locally, they show a spaced foliation, more evident towards the contact with the country schists, belonging to the "*Filladi inferiori*" Fm; where present, foliation wraps the tourmaline spots (Fig. 4a). The contacts with the country rocks are usually tectonized and are locally characterized by the occurrence of breccias and/or by the comminution of the grain size, resulting in a blackening of the rock, as observed, for instance, along the border of the rock body occurring at Le Casette (Fig. 3e). Moreover, around the Fornovolasco Metarhyolite rock bodies, the metagreywackes and phyllites belonging to the "*Filladi inferiori*" Fm are distinctly tourmalinized and tourmaline veins, up to 1 cm in thickness, cut the Fornovolasco Metarhyolite Fm itself (Fig. 3d). Usually, in the tourmalinized outcrops, the occurrence of thallium-bearing pyrite veins, associated with minor arsenopyrite and trace amounts of other sulfides and sulfosalts (e.g., galena, tetrahedrite) has been observed (e.g., in the Boscaccio area – Fig. 3f).

The largest outcrops of this formation occur in the Fornovolasco tectonic window, whereas small and scattered outcrops are present in the Stazzemese area and in the Sant'Anna tectonic window (Fig. 2). As reported by Pieruccioni et al. (2018), the largest rock body is exposed close to Fornovolasco, in the Boscaccio locality, where it forms a lensoidal body of ca. 1.5 km in length and 120 m in apparent thickness. Another significant outcrop was described by Bonatti (1933), close to Le Casette locality.

The bodies of these porphyritic rocks are usually flattened on the main field composite foliation. In the Fornovolasco area, this foliation records Alpine D1 and D2 events as well as pre-Alpine deformation (Variscan in age), as widely reported elsewhere in the Paleozoic basement of Alpi Apuane (e.g., Conti et al., 1991). Only locally, e.g., at Le Casette and Boscaccio, porphyritic rock bodies can be observed discordantly cutting the foliation (Variscan ?), which is crenulated and transposed in the Alpine main foliation far from the porphyritic body (Fig. 5b).

*4.2. Petrography and mineral chemistry*

The Fornovolasco metarhyolite is characterized by a porphyritic texture (Figg. 4b-c, 6), with quartz, mica, and feldspar phenocrysts set in a fine-grained quartz-white mica-feldspar groundmass. Some samples are foliated, with porphyroclasts of quartz, tourmaline, and feldspar, wrapped by the foliation (Fig. 6c). Tectonic deformation may result in a reduction of the grain size, as observed in the blackened variety of metarhyolite found in the marginal portions of the outcrop of Le Casette.

Quartz is usually represented by anhedral to subhedral grains, up to 5 mm in size, sometimes showing rounded embayment structures (Fig. 6d). Euhedral hexagonal crystals sections have been also observed (Fig. 6e). Usually, such crystals display the formation of subgrain boundaries and undulose extinction, indicating their metamorphic recrystallization.

Feldspars occur both as albite (containing less than 0.01 K atoms per formula unit) and K-feldspar ($Or_{95}Ab_5 - Or_{99}Ab_1$). Usually, feldspar phenocrysts are completely replaced by fine-grained white mica (Fig. 6a). Only rarely, anhedral grains of K-feldspar have been identified at the core of aggregates of fine-grained white mica. Albite occurs both as phenocrysts, likely replacing magmatic plagioclase (Fig. 6f), showing Albite and Albite-Carlsbad twins, or as µm-sized anhedral twinned grains (Fig. 6c), flattened on the main field composite foliation (when present). Probably, this second type of albite could be related to the metamorphic recrystallization of the metarhyolite.

Phyllosilicates are represented by biotite, chlorite, and white mica (Table 2). Biotite is a relict igneous mineral and it is only rarely preserved (Fig. 6g), usually being replaced by an assemblage made of chlorite + quartz ± rutile, sometimes with inclusions of zircon, apatite, and ilmenite. Taking into account its average chemical composition, $K_{0.89}(Fe_{1.53}Mg_{1.00}Ti_{0.16}Al_{0.15}V_{0.02})_{\Sigma2.86}(Al_{1.14}Si_{2.86})O_{10}(OH_{1.97}F_{0.03})$, it can be classified as a Mg-rich annite (Rieder et al., 1998; Tischendorf et al., 2007). Chlorite, occurring in samples FVb7 (after biotite) and SAS1 (both after biotite and as synkinematic porphyroblasts aligned on the main field composite schistosity), has an average chemical composition $(Mg_{2.45}Fe_{2.22}Mn_{0.02}Al_{1.30}Ti_{0.01})(Si_{2.70}Al_{1.30})O_{10}(OH)_8$, corresponding to a Fe-rich clinochlore (Wiewióra and Weiss, 1990).

White mica usually occurs as subhedral crystals showing a decussate texture; it can also form fine-grained individuals, closely associated with quartz and minor feldspar. On the basis of its chemistry, it can be classified as muscovite (Rieder et al., 1998; Tischendorf et al., 2007). Back-scattered electron images show a distinct chemical zoning, with dark and bright regions. This zoning is mainly related to the phengitic substitution. Only a minor paragonitic component was observed.

Tourmaline is widespread in almost all the studied samples, even if its modal abundance is highly variable. Tourmaline occurs as euhedral to anhedral grains, scattered within the groundmass, or as rounded spots, formed by the intimate intergrowth of tourmaline, quartz, and minor albite. In some cases, radiated aggregates of prismatic to acicular tourmaline crystals have been observed (Fig. 7). In the more deformed samples, tourmaline crystals are strongly fractured, with fractures filled by quartz, and the tourmaline + quartz ± albite spots are wrapped by foliation. Optically, tourmaline crystals show a very complex zoning (Figg. 6h, 7), with colors ranging from colorless to green-bluish to dark brown. No systematic chromatic trends between cores and rims could be observed. However, it should be stressed that a clear distinction between the different portions of tourmaline crystals can be difficult, since this mineral usually forms polycrystalline aggregates. Rarely, a clear identification between cores and rims was possible. In these cases, cores usually show a bluish color, with brown to dark brown rims, whereas an inverse color pattern (brownish cores and bluish rims) was rarely observed.

This optical zoning is related to a chemical zoning, patchy or oscillatory, revealed through SEM observations. Electron microprobe analyses on tourmaline from the Fornovolasco Metarhyolite were performed on samples FVb7 (showing the best-preserved magmatic textures, with biotite relicts and tourmaline spots) and FVb9, with tourmaline crystals scattered in the groundmass or forming radial fibrous aggregates. Results are reported as Supplementary Material SM1, whereas selected chemical analyses of tourmaline crystals, representative of the observed chemical variations discussed below, are given in Table 3. Cation proportions of tourmaline from the Fornovolasco metarhyolite are plotted in the Al–Fe(tot)–Mg and Ca–Fe(tot)–Mg ternary diagrams of Henry and Guidotti (1985) (Fig. 8). Following the nomenclature of Henry et al. (2011), tourmalines occurring in the studied samples are members of the dravite-schorl series. Since accurate determinations of B, O, H, Li are lacking, and the oxidation states of transition elements are unknown, the chemical formulae were recalculated on the basis of 31 anions, assuming the occurrence of 3 B and 4 OH groups per formula unit (i.e., on the basis of 24.5 O atoms, in agreement with Henry et al., 2011). The members of the tourmaline supergroup are characterized by the general structural formula $XY_3T_6Z_6O_{18}(BO_3)_3V_3W$ (Henry et al., 2011), where italic letters indicate different sites. Tourmaline from the Fornovolasco metarhyolite belongs to the alkali group, having Na as the dominant $X$ cation. The total number of $X$ cations ranges between 0.60 and 0.99 atoms per formula unit. The occurrence of vacancies is related to the alkali-defect substitution $^{X}Na^{+} + {}^{Y}(Mg, Fe)^{2+} = {}^{X}\square + {}^{Y}Al^{3+}$. Minor Ca (up to 0.32 apfu) replaces Na, according to the uvite substitution $^{X}Na^{+} + {}^{Z}Al^{3+} = {}^{X}Ca^{2+} + {}^{Z}(Mg, Fe)^{2+}$. No significant differences in the $X$ cation contents between zones having different colors were identified: bluish zones have a total $X$ content ranging from 0.60 to 1.07 apfu, whereas brownish area show an $X$ content variation between 0.63 and 0.99 apfu.

Octahedral $Y$ cations are mainly represented by Mg and Fe, with Mg/(Mg + Fe) atomic ratios ranging between 0.30 and 0.96. As shown by Fig. 8, the Mg/(Mg+Fe) ratio in tourmalines occurring in the sample FVb7 is quite constant [average = 0.50(4); range = 0.44-0.58], whereas it is very variable in sample FVb9, ranging from 0.30 to 0.96, with an average Mg/(Mg + Fe) atomic ratio of 0.71(19). Magnesium and Fe can be replaced by Al, up to ~ 0.80 apfu. Additionally, Al is the dominant Z. In some cases, dark brown zones of tourmaline from sample FVb7 show up to ca. 1 Mg apfu. Tourmaline crystals from sample FVb9 are very rich in Al, in some cases showing more than 6.5 Al apfu, indicating not only the occurrence of this element at the Z site but also the occurrence of not negligible $^{Y}Al$.

The role of Ti in tourmaline is not completely constrained yet. According to Henry et al. (2011), this element should be considered a $Y$ cation, whereas Novák et al. (2011) suggested its occurrence at the $Z$ site. The amount of Ti seems to be related to the color shown in plane-polarized light by tourmaline crystals. Indeed, as shown in Figure 7, the brown to dark brown colors are characterized by the highest Ti contents. This is particularly evident in tourmaline from sample FVb7, where dark brown zones have Ti contents up to 0.57 apfu. Also, there is a positive correlation between Ti and the sum (Mg+Fe) and a negative one with the Al content. This could suggest suggest the substitution mechanism $(Mg,Fe)^{2+} + 2Al^{3+} = Ti^{4+} + 2(Mg,Fe)^{2+}$. The total number of Y cations falls in the range 2.71 – 2.98 apfu.

The *T* cations are represented by Si (with average values close to 5.90 apfu) and minor Al. As stated above, owing to the lack of complete analyses of tourmaline, the occurrence of 4 OH groups (at *V* and *W* sites) was assumed. Whereas the occurrence of F can be discarded, this element being below the detection limit, the presence of minor $O^{2-}$ replacing $OH^-$ at the *W* site cannot be excluded.

Accessory minerals are represented by apatite, epidote, monazite-(Ce) (in some cases included in apatite), rutile (usually after biotite), titanite, and zircon. Minor amounts of sulfides (pyrite, arsenopyrite, and traces of galena and tetrahedrite) have been observed in some samples.

*4.3. Geochemistry*

The studied rocks plot into the rhyolite field of the Total Alkali vs Silica classification diagram (Le Bas et al., 1986; Fig. 9a) and into the dacite/rhyodacite and trachyandesite fields of the $Zr/TiO_2$ vs Nb/Y diagram (Winchester and Floyd 1977; Fig. 9b). Having to choose between metarhyolite and metadacite/metarhyodacite/metatrachyandesite, we opted for metarhyolite to emphasize the $SiO_2$- and quartz phenocryst-rich nature of studied rocks. The Fornovolasco metarhyolite is peraluminous having an Alumina Saturation Index (ASI) variable between 1.3 and 3.2. The trace-element signature of these rocks is that typical of highly evolved orogenic magmas showing negative anomalies of Nb, Ta, P, and Ti in the Primitive Mantle-normalized patterns (Tab. 4, Fig. 10a). The relative concentrations of Nb, Y, and Zr, that should be close to the original ones, even following moderate alteration and low-grade metamorphism (e.g., Pearce et al., 1984), are also indicative of an orogenic signature. The low contents of Sr and Ba, coupled with the negative Eu anomalies (Eu/Eu* = 0.52-0.85; Fig. 10a, b) suggest extensive feldspar fractionation.

With respect to the felsic metavolcanic rocks (metarhyolite and metadacite; Fig. 9a) of the "*Porfiroidi e scisti porfirici*" Fm, the Fornovolasco metarhyolite has similar incompatible trace element distribution, systematically lower Zr, Hf, Y, Nb, Ba, Th and REE, and slightly higher

LREE/HREE ratios (Fig. 10a, b). The moderate geochemical differences between the two rock suites could be either primary (magmatic) or due to the metamorphic/hydrothermal processes suffered by the rocks. It is worth noting that some samples from Fornovolasco (e.g., FVb1, FVb13, and FVc2) show high enrichments in Pb, Bi, As, and Sb (Fig. 10a, Tab. 4).

*4.4. U-Pb zircon age*

The LA-ICP-MS U-Pb analyses were performed on the four selected samples, representative of the main outcrops of the Fornovolasco Metarhyolite Fm in southern Alpi Apuane. Selected zircon crystals are generally euhedral and range in length between 100 and 300 μm. Most of them have an elongated prismatic habit, with a width-to-length ratio > 2 (~ 81%), even if equant crystals occur (~ 19%). Some crystals contain fluid inclusions or mineral grains. The SEM-CL images (Fig. 11) revealed a complex internal structure. Usually, zircon crystals show a well-developed oscillatory zoning. About 28% of zircon crystals show structureless xenocrystic cores, even if some cores characterized by oscillatory or sector zoning have been observed; in some cases, faint oscillatory zoning can be distinguished. The following results have been obtained (all errors are given at the 2σ level):

**FVb13** – A total of 116 zircons were chosen. Among them, 90 have an elongated prismatic habit. SEM-CL images show the presence of 32 cores. A total of 122 LA spots on 82 zircons gave 46 (36 rims + 10 cores) U-Pb concordant ages, varying between 242 ± 4 to 301 ± 5 Ma ($^{206}Pb/^{238}U$ ages). Core ages varied from 260 ± 7 Ma to 298 ± 4 Ma. Weighted average ages for rims and cores are 271 ± 4 and 277 ± 9 Ma, respectively.

**FVc1** – A total of 80 zircons was selected. Most of them show an elongated prismatic habit (61). SEM-CL images show the occurrence of 22 cores. A total of 85 LA spots on 61 zircons gave 47 (40 rims + 7 cores) U-Pb concordant ages, with rim and core ages varying from 256 ± 6 to 296 ±

4 Ma and from 270 ± 6 to 294 ± 6 Ma ($^{206}$Pb/$^{238}$U ages), respectively. Weighted average ages range between 280 ± 3 Ma (rims) and 283 ± 7 Ma (cores).

**POL1** − A total of 107 zircons were selected; 85 of them have an elongated prismatic habit. SEM-CL images show the occurrence of 33 cores. A total of 73 LA spots were performed on 55 zircons, giving 24 U-Pb concordant ages (only on rims) varying between 226 ± 6 and 306 ± 6 Ma ($^{206}$Pb/$^{238}$U ages). The weighted average age is 277 ± 8 Ma.

**SAS1** − A total of 114 zircons were chosen, 97 showing an elongated prismatic habit. SEM-CL images show the presence of 29 cores. A total of 106 LA spots, performed on 80 zircons, gave 52 (44 rims + 8 cores) U-Pb concordant ages, varying between 259 ± 7 to 311 ± 5 Ma ($^{206}$Pb/$^{238}$U ages), with core ages varying from 281 ± 5 Ma to 301 ± 5 Ma. Weighted average ages on cores and rims are 290 ± 6 and 292 ± 3 Ma, respectively.

Table 5 and Figures 12 and 13 summarize such results. The complete U-Pb data set is reported as Supplementary Material SM2. A total of 386 LA spots have been collected on 278 different zircon crystals of the Fornovolasco metarhyolite. They gave 169 U-Pb concordant ages (Fig. 12), having a weighted average age of 280 ± 2 Ma. However, the probability density plot estimation reported in Figure 12 does not show a simple Gaussian distribution, suggesting the contribution of at least two different populations, plus few occurrences at younger ages. The main population could resolve an event occurring at the modal value (~ 288 Ma), whereas a second event could be represented by relatively younger ages (at ~ 270 Ma).

Figure 13 shows the different age distributions for the four studied samples. Samples FVc1 and SAS1 have the higher percentage of concordant data (55% and 49%, respectively), whereas FVb13 and POL1 have only 38% and 33% of concordant data, respectively. The 217 discordant ages were excluded. Among the U-Pb concordant ages, 164 data indicate an age spread over a 50 Ma period, from late Carboniferous to middle Permian (Figure 13). The remaining five concordant data, belonging to samples FVb13 and POL1, gave younger ages, ranging from 226 ± 6 to 244 ± 9 Ma.

## 5. Discussion

*5.1. The Fornovolasco metarhyolite: a Permian magmatic product*

Previous authors debated the nature of the porphyritic tourmaline-rich rocks of the southern Alpi Apuane, assigning them different origins (e.g., Bonatti, 1933). Pandeli et al. (2004), on the basis of petrographic features, described such rocks as metavolcanic products. The textural and mineralogical features observed in all the studied samples (Table 1), representative of the different outcrops of the Fornovolasco Metarhyolite Fm, support their magmatic origin. Indeed, they show a porphyritic texture, with euhedral phenocrysts of feldspar and quartz, the latter with embayment structures, coupled with the occurrence of relict femic phases (e.g., biotite), tourmaline spots, and zircon crystals showing magmatic *habitus* (Figg. 3, 4, 6, 11).

The Alpine tectono-metamorphic overprint partially obliterated the original petrographic features of these rocks. In addition, the widespread tourmalinization and the local occurrence of pyrite veins (Fig. 3d, f) suggest that these rocks had been affected by hydrothermal processes with an additional possible alteration of the original compositional and textural characteristics. As a matter of fact, it is well-known that the selective alteration and deformation of felsic igneous rocks in hydrothermal-metamorphic conditions could result in rocks mainly formed by quartz and white mica (e.g., Etheridge and Vernon, 1981; Williams and Burr, 1994; Vernon, 2004), with only a few relicts of the more alterable phases (e.g., feldspar, biotite). The Fornovolasco metarhyolite shows variable textures, from well-preserved porphyritic ones, with relict biotite, feldspar, and embayed magmatic quartz still preserved (Fig. 6), to strongly deformed rocks, where the magmatic texture was partially to completely obliterated.

Notwithstanding such alteration and deformation processes, the structural as well as the petrographic data support a subvolcanic setting for the Fornovolasco metarhyolite. For instance, the study of the field relationships between some metarhyolite bodies and the foliation of the surrounding schists (Fig. 5) suggests that these rocks originally emplaced as dykes or laccoliths. A subvolcanic nature is also supported by the occurrence of tourmaline spots and agrees with the textural features of the Fornovolasco Metarhyolite (e.g., porphyritic texture, embayed quartz – Figg. 3, 6). Indeed, tourmaline spots are a distinctive feature mostly occurring in leucocratic intrusive rocks (e.g., Buriánek and Novák, 2007; Perugini and Poli, 2007; Hong et al., 2017 and references therein); even if their origin is still debated, they are usually related to an intrusive or subvolcanic magmatic-hydrothermal setting. The formation of the tourmaline spots clearly predates the Alpine tectono-metamorphic events, being wrapped by the Alpine foliation (when the latter is developed within strongly deformed horizons in the metarhyolite rock bodies – Fig. 4a).

As stated above, the U-Pb concordant ages extend over a time interval of about 50 Ma, from late Carboniferous to middle Permian. Such a spread is common to the four studied samples, although every one displays a different age distribution within this interval (Fig. 13). It is worth noting that such a variability is common to several Permian magmatic rocks throughout Europe (e.g., Cortesogno et al., 1998; Finger et al., 2003; Visonà et al., 2007; Dallagiovanna et al., 2009). Owing to the general homogeneity of the petrographical and geochemical features of the Fornovolasco metarhyolite occurring over the studied area, a single magmatic event could be hypothesized. Taking into account the U-Pb ages and their weighted averages (Table 5), an early-middle Permian age can be proposed, fitting with the regional framework (Fig. 14) showing the occurrence of sedimentary basins, as testified by the *Scisti di San Lorenzo* and *Rio Marina* Fms (dated at 280 Ma – Siirevaag et al., 2016; Paoli et al., 2017), as well as a thermal event, recorded in the *Filladi di Buti* Fm (Monti Pisani) and in the *Micascisti* Fm in the Larderello sub-surface and dated between 275 and 285 Ma (Borsi et al., 1966; Del Moro et al., 1982; Ferrara and Tonarini, 1985). Moreover, it agrees with the ages of several peraluminous felsic and minor mafic magmatic bodies emplaced during the post-Variscan extensional phase in the southern sector of the Variscan belt in Western Europe (e.g., Stampfli and Borel, 2002; Stampfli et al., 2002; Schuster and Stüwe, 2008; von Raumer et al., 2016). The probability density distribution of zircon ages showing two different zircon populations reported in Fig. 12 could be interpreted as the result of an early magmatic crystallization (at ~ 288 Ma), corresponding to the modal value, and a subsequent post-magmatic event affecting the original U-Pb zircon systematics leading to younger ages (e.g., Bomparola et al., 2007). The nature of this event is presently not known, even if it could be hypothetically related to hydrothermal alteration and sulfide deposition, occurring in some of these rock bodies and geochemically expressed as high contents of As, Sb, Pb, and Bi (Fig. 10, Table 4). The second population of ages is particularly frequent in samples FVb13 and POL1, having zircon showing younger ages and the lower percentage of concordant data. Such younger ages agree with the results obtained by Molli et al. (2002) on amphiboles from the Cerreto metamorphic slices (northern Alpi Apuane), that showed a late Permian – early Triassic overprint (~ 240 Ma) possibly related by these authors to the Variscan orogenic collapse or the Tethyan rift-related crustal thinning.

An alternative interpretation of the time spread shown by the data (Figg. 12, 13) could be that the older concordant ages, ranging between 275 and 310 Ma, measured on xenocrystic and oscillatory-zoned cores and rims, are due to inherited components from the source rocks, whereas the age of the magmatic event could range between 260 and 275 Ma, as resulting from the youngest $^{206}Pb/^{238}U$ ages measured on oscillatory zoned rims. This interpretation agrees with the results of Dallagiovanna et al. (2009), who found inherited cores only slightly older than the magmatic rims.

Further studies are needed to propose a clear picture of the Permian magmatism recorded in the Alpi Apuane Metamorphic Complex. As a matter of fact, the Fornovolasco metarhyolite represents the record of a magmatic event following the Variscan orogeny and gives new hints on the Permian evolution of the basement of the Adria plate.

*5.2. The Variscan magmatism in Northern Apennines*

The occurrence of a late- to post-Variscan magmatism in the Northern Apennines was hypothesized by several authors through the identification of metavolcanoclastic Permian formations in some of the small and scattered outcrops of Paleozoic rocks occurring in Tuscany (Fig. 14). For instance, the "*Scisti porfirici di Iano*" Fm (central Tuscany) is made of rhyolitic metavolcaniclastic rocks (e.g., Barberi, 1966; Lazzarotto et al., 2003 and references therein), and abundant fragments of volcanic rocks have been found in the "*Arenarie rosse di Castelnuovo*" Fm, known in the Larderello subsurface (southern Tuscany; Bagnoli et al., 1979) and in the AGIP "Pontremoli 1" well (northern Tuscany; Pandeli et al., 1994). It is worth noting that the Permian age attribution is only based on stratigraphic correlations. Moreover, pebbles of red rhyolitic porphyries, similar to the Permian rhyolites from southern Alps and Sardinia, were found in the Middle Triassic Verruca Fm of the Monti Pisani (northern Tuscany) (e.g., Puxeddu et al., 1984). Finally, several megaliths of granitoid rocks occurring in the External Liguride units have been interpreted as fragments of the continental crust partially derived from the Adria plate margin. Their ages, obtained by Eberhardt et al. (1962) through K/Ar and Rb/Sr dating on muscovite, range between 272 ± 16 and 310 ± 10 Ma.

The age of the Variscan and post-Variscan metamorphism in the basement of the Northern Apennines was dated by several authors. Molli et al. (2002) reported $^{40}Ar/^{39}Ar$ ages measured on amphibole of 328 – 312 Ma for the MORB-derived ortho-amphibolites of the Cerreto metamorphic slices, whereas Lo Pò et al. (2016), on the basis of the chemical dating of monazite, suggested a metamorphic peak at 292.9 ± 3 Ma for the metasedimentary rocks recovered in the "Pontremoli 1" well. The occurrence of a post-Variscan high-*T* thermal event was previously suggested by Del Moro et al. (1982), on the basis of a Rb/Sr age of 285 ± 11 Ma obtained for muscovite associated with andalusite in the "*Micascisti*" Fm in the Larderello subsurface. Such an age is similar to those found using Rb/Sr dating on whole rock for the "*Filladi e Quarziti di Buti*" Fm (Monti Pisani), i.e., 285 ± 12 Ma (Borsi et al., 1966; Ferrara and Tonarini, 1985). Consequently, this formation, recently dated through U-Pb zircon dating to the late Proterozoic – early Cambrian by Paoli et al. (2017), possibly recorded also an early-middle Permian metamorphism.

The thermal events recorded in the Paleozoic basement could be related to the thinning of the lithosphere and the emplacement of magmatic bodies following the Variscan orogeny, e.g., gabbro-derived granulites from the External Liguride Units having Sm/Nd ages of ~ 290 Ma obtained on clinopyroxene, plagioclase, and whole rock (Marroni and Tribuzio, 1996; Meli et al., 1996; Marroni et al., 1998; Montanini and Tribuzio, 2001). This evidence agrees with the dating of the Fornovolasco metarhyolite, that can be considered as the first evidence of Permian subvolcanic magmatic bodies intruded in the Paleozoic basement of the Northern Apennines.

*5.3. Permian magmatism and ore genesis in the Alpi Apuane*

As pointed above, the Fornovolasco metarhyolite is spatially associated with the main ore deposits from the southern Alpi Apuane (Fig. 2), as well as to tourmalinite bodies. Moreover, the widespread occurrence of tourmaline in the metarhyolite is a distinctive feature (Fig. 3).

The occurrence of tourmalinite in the Paleozoic basement of the southern Alpi Apuane has been known since a long time and was used by miners as an empirical prospecting tool (e.g., Benvenuti et al., 1989). Several origins have been proposed for such a rock: magmatic (D'Achiardi, 1885), magmatic-hydrothermal, related to a hypothetical Tertiary magmatism (Carmignani et al., 1972, 1975), and sedimentary-exhalative (Benvenuti et al., 1989). According to these latter authors, the occurrence of tourmalinite could indicate a pre-Alpine concentration of boron and metals, later remobilized during the Alpine tectono-metamorphic events. Indeed, several authors proposed a genetic link between tourmalinite and ore deposits (e.g., Lattanzi et al., 1994 and references therein). As a consequence, the age of the tourmalinization event has been taken as the age of the proto-ores. Following Benvenuti et al. (1989), an early Paleozoic age for the tourmalinization process was suggested, related to the middle Ordovician "*Porfiroidi*" magmatic cycle, even if such an age was attributed only to the Pb-Zn-Ag ore bodies. On the contrary, a possible Permo-Triassic age was suggested for the pyrite ± baryte ± iron oxide ore deposits and the associated tourmalinite (e.g., Costagliola et al., 1990). Finally, D'Orazio et al. (2017), based on their identical Pb-isotope composition, proposed a common early Paleozoic origin for all the major Pb-Zn-Ag and pyrite ± baryte ± iron oxide deposits from the southern Alpi Apuane.

The discovery of a Permian magmatism in this area allows refining this minerogenetic model. Indeed, tourmalinization seems to be related to the occurrence of the metarhyolite bodies, in agreement with the usual association of tourmaline + quartz hydrothermal veins and metasomatic bodies with shallow crustal level tourmaline-bearing felsic rocks (e.g., Dini et al., 2008). Moreover, lead isotope composition obtained for sulfides and sulfosalts from the ore deposits of southern Alpi Apuane is similar to that shown by the ores generated during the Paleozoic metallogenetic event of Sardinia and its Variscan magmatic rocks (D'Orazio et al., 2017). The lack of knowledge about the occurrence of a Permian magmatism led these authors to hypothesize a genetic link with the middle Ordovician magmatism. On the contrary, the present study allows discarding this hypothesis, suggesting a Permian age for the major ore deposits from southern Alpi Apuane. It is worth noting that the geochemistry of these pyrite ores is typical of low-$T$ hydrothermal systems, in some cases associated with a coeval felsic magmatism occurring in an evolved continental crust (e.g., Lentz, 1998). It is very likely that the study area was characterized by a similar geodynamic setting in post-Variscan time, in agreement with several authors (e.g., Rau, 1990; Marroni et al., 1998; Padovano et al., 2012; Lo Pò et al., 2016).

A pre-Ladinian tourmalinization event was suggested by Cavarretta et al. (1989, 1992), owing to the occurrence of tourmalinite clasts within the Middle Triassic metasiliciclastic Verrucano sequence. These authors hypothesized a Permian high-$T$ hydrothermal activity, associated with acid magmatism, related to the extensional phase of the Variscan orogeny. Moreover, on the basis of the crystal chemistry of tourmaline in the tourmalinite clasts, Cavarretta et al. (1989, 1992) proposed an exhalative origin for tourmaline, with the interaction between magmatic fluids and Mg-rich cold seawater. A similar hypothesis was put forward by Benvenuti et al. (1991) for tourmaline from the Bottino mine (southern Alpi Apuane).

Petrographic observations and chemical data show the complex zoning of tourmaline from the metarhyolite bodies. Color zoning seems to be mainly related to the Ti content, in agreement with previous authors, with dark brown areas characterized by the higher contents of this element (e.g., Plimer and Lees, 1988; Slack and Coad, 1989; Benvenuti et al., 1991). As shown in Figure 8, tourmalines from sample FVb7 have a chemical composition similar to those described by Benvenuti et al. (1991) from the tourmalinite bodies of the Bottino mining district. On the contrary, tourmaline occurring in sample FVb9, scattered in the groundmass, has a wide compositional range, from Fe-rich to Mg-rich composition.

The close similarity between the chemistry of tourmaline from metarhyolite bodies and tourmalinite is also marked by the occurrence of *Y* vacancy, as reported also by Benvenuti et al. (1991). This vacancy may be interpreted in two different ways, owing to the lack of a complete analysis of tourmaline. The vacancy could be apparent, and related to a small oxy-component, with minor (OH)$^-$ replaced by O$^{2-}$, or it could be real, in agreement with some structural refinements showing small numbers of vacancies at both *Y* and *Z* sites (e.g., Ertl et al., 2003; Prowatke et al., 2003). The former hypothesis, proposed by Benvenuti et al. (1991) to explain the deficit of octahedral cations observed in tourmaline from the Bottino district, could also be supported by the relatively high Ti content of some grains, favoring the occurrence of an oxy-component. It is worth noting that the dark brown areas showed Ti contents among the highest ever recorded in natural samples (e.g., Scribner et al., 2018). Usually, high-Ti tourmaline can also have Fe$^{3+}$ (e.g., Bosi et al., 2017). However, data about the oxidation state of Fe in the studied samples is currently not available. Further studies on the crystal chemistry of this tourmaline are needed to address such an issue. The close similarity between the chemistry of tourmaline from metarhyolite bodies and tourmalinite could suggest their common origin, likely related to hydrothermal circulation.

It is well-known that no specific compositional field can be established in the diagrams of Henry and Guidotti (1985) for tourmaline from tourmalinites, owing to their wide compositional range (e.g., Plimer and Lees, 1988; Slack, 1996; Torres-Ruiz et al., 2003). Indeed, tourmaline from the Fornovolasco metarhyolite falls in the compositional fields corresponding to clastic metasediments, Li-poor granitoids, and $Fe^{3+}$-rich quartz-tourmaline rocks, in agreement with occurrences from other localities worldwide (e.g., tourmaline in tourmalinites from the Betic Cordillera, Spain – Torres-Ruiz et al., 2003). Consequently, the diagram of Henry and Guidotti (1985) can be used only to illustrate the wide chemical variability of tourmaline from the Fornovolasco metarhyolite, avoiding any petrogenetic implication. It is worth noting that such compositional variations likely reflect the pre-metamorphic magmatic-hydrothermal as well as syn- to post-metamorphic processes affecting the basement of the Alpi Apuane, and constitute an essential tile in the reconstruction of the genesis of ore deposits (e.g., Slack, 1996 and references therein).

Consequently, a more detailed study of the chemical and isotopic composition of tourmaline seems to be mandatory.

## 6. Summary and conclusion

The data reported in this work document that the Fornovolasco metarhyolite is the first occurrence of Permian felsic magmatic rocks within the basement of Northern Apennines, where only metavolcanoclastic rocks were known in Permo-Triassic metasediments. This finding fits with the previous knowledge in other sectors of the West-Mediterranean southern Europe, such as Eastern and Southern Alps, Ligurian Alps; Corsica, Sardinia, and Calabria (e.g., Dal Piaz, 1993; Cortesogno et al., 1998; Visonà et al., 2007; Schuster and Stüwe, 2008; Dallagiovanna et al., 2009; Festa et al., 2010; Oggiano et al., 2010; Zibra et al., 2010). The Fornovolasco metarhyolite shares its calc-alkaline nature with the late Carboniferous to early Permian magmatic products occurring in these areas and referred to a generalized lithospheric thinning associated with a regional-scale crustal wrenching (e.g., Deroin and Bonin, 2003; Dallagiovanna et al., 2009 and references therein). The discovery of Permian magmatic rocks within the basement of the Northern Apennines gives new hints for a better knowledge and understanding of the geodynamic evolution of the "Apenninic" portions of the Adria plate basement. Moreover, it fits the results of previous studies on the Northern Apennine basement, showing the occurrence of a post-Variscan thermal event recorded in the basement rocks at Permian time, hypothetically related to the thinning of the lithosphere and the emplacement of magmatic bodies. Finally, the finding of Permian magmatic rocks in close spatial association with the ore deposits of the Alpi Apuane Metamorphic Complex represents a significant constraint on the origin and age of these ores.

**Acknowledgements**

This research received support by Ministero dell'Istruzione, dell'Università e della Ricerca through the project SIR 2014 "THALMIGEN – Thallium: Mineralogy, Geochemistry, and Environmental Hazards" (Grant No. RBSI14A1CV), granted to CB. MP wish to acknowledge the European Research Council for the Consolidator grant CHRONOS (Grant No. 612776 – P.I. Diego Perugini). The constructive criticism of two anonymous reviewers helped us improving the paper.

**Table Captions**

**Table 1**. Details of the metarhyolite rock samples from Alpi Apuane studied in this work.

**Table 2**. Chemical analyses of phyllosilicates occurring in the Fornovolasco metarhyolite.

**Table 3**. Selected chemical analyses of tourmaline occurring in the Fornovolasco metarhyolite.

**Table 4**. Major- and trace-element analyses of metarhyolite rock samples from Alpi Apuane studied in this work.

**Table 5**. Summary of LA-ICP-MS U-Pb zircon datings of the Fornovolasco metarhyolite.

**Figure Captions**

**Fig. 1**. Geological sketch map of the Alpi Apuane Metamorphic Complex (based on Carmignani and Kligfield, 1990 and modified after Molli and Vaselli, 2006). The studied area is shown in the boxed frame. Red circles indicate the sampling sites of "*Porfiroidi*" (1 = PORF1; 2 = PORF2; 3 = PORF3 – see Table 1).

**Fig. 2**. Simplified geological map of the southern Alpi Apuane showing the location of the outcrops of the Fornovolasco Metarhyolite Fm (same labels as in Table 1) and the main ore deposits: 1) Fornovolasco; 2) La Tana; 3) Buca della Vena; 4) Canale della Radice; 5) Bottino; 6) Monte Rocca; 7) Argentiera di Sant'Anna; 8) Monte Arsiccio; 9) Buca dell'Angina; 10) Pollone; 11) Levigliani. The outcrop close to Le Casette (boxed area) is shown in detail in Figure 5.

**Fig. 3**. Field features of the Fornovolasco Metarhyolite Fm. a) Hand specimen showing the occurrence of cm-size black tourmaline spots (base of the triangle = 12 cm). Boscaccio, Fornovolasco; b) the massive and granular nature of the metarhyolite cropping out at Le Casette, Fornovolasco (detail in Fig. 4d); c) massive and porphyritic metarhyolite, with tourmaline spots, cropping out at Trimpello, Fornovolasco (details in Fig. 4c and 6); d) tourmaline veins and spots in the metarhyolite body cropping out at Boscaccio, Fornovolasco; e) the black facies occurring at Le

Casette, Fornovolasco; f) Tl-rich pyrite veins embedded in the metarhyolite, with black tourmaline spots. Boscaccio, Fornovolasco.

**Fig. 4**. Polished slabs of specimens of the Fornovolasco Metarhyolite, showing some textural details. a) Foliated metarhyolite, with tourmaline spots wrapped by the main foliation. Boscaccio, Fornovolasco; b) porphyritic metarhyolite with quartz phenocrysts from Sant'Anna di Stazzema; c) porphyritic metarhyolite, with quartz and feldspar phenocrysts, from Trimpello, Fornovolasco; d) granular metarhyolite, with abundant black tourmaline, from Le Casette, Fornovolasco. Labels as in Table 1.

**Fig. 5**. (a) Geological map and cross-section showing the outcrops of the Fornovolasco Metarhyolite Fm at Le Casette, Fornovolasco. In the cross-section, the thin red lines indicate the relics of Variscan foliation. (b) Detail of the contact (red line) between the "*Filladi inferiori*" Fm (FAF), showing a pervasive foliation (light blue lines), and the massive rocks belonging to the Fornovolasco Metarhyolite Fm (FMR). Boscaccio, Fornovolasco.

**Fig. 6**. Petrographic features of the Fornovolasco Metarhyolite. a) Porphyritic texture, showing phenocrysts of quartz and sericitized feldspars. b) Quartz phenocrysts, with euhedral to subhedral morphology, in a sericitized groundmass. c) Tourmaline spot, formed by the intimate association of tourmaline + quartz, wrapped by the Alpine schistosity. d) Quartz phenocrysts with magmatic embayment, in a sericitized groundmass. e) Euhedral quartz phenocryst. f) Euhedral crystal of twinned feldspar, with an incipient crystallization of white mica. g) Biotite flakes, partially replaced by chlorite. h) Subhedral crystals of tourmaline showing a distinct zoning. Abbreviations: Ab = albite; Ab-Pl = albite after plagioclase; Bt, biotite; Chl, chlorite; Qz, quartz; Tur, tourmaline; Fld psd, feldspar pseudomorph.

**Fig. 7**. Variation in cations (atoms per formula unit, apfu) across selected tourmaline crystals from samples FVb7 and FVb9.

**Fig. 8**. Chemical composition of tourmaline from the Fornovolasco metarhyolite plotted in the Al-Fe(tot)-Mg (a) and Ca-Fe(tot)-Mg (b) diagrams. Symbols: squares = FVb7; circles = FVb9; black crosses = tourmalinite (Benvenuti et al., 1991); black stars = "*Filladi inferiori*" Fm (Benvenuti et al., 1991); and white stars = "*Porfiroidi e Scisti porfirici*" Fm (Benvenuti et al., 1991). Compositional fields after Henry and Guidotti (1985). In (a): (1) Li-rich granitoid pegmatites and aplites; (2) Li-poor granitoids and their associated pegmatites and aplites; (3) $Fe^{3+}$-rich quartz-tourmaline rocks (hydrothermally altered granites); (4) metapelites and metapsammites coexisting with an Al-saturating phase; (5) metapelites and metapsammites not coexisting with an Al-saturating phase; (6) $Fe^{3+}$-rich quartz-tourmaline rocks, calc-silicate rocks, and metapelites; (7) low-Ca metaultramafics and Cr,V-rich metasediments; (8) metacarbonates and meta-pyroxenites. In (b): (1) Li-rich granitoid pegmatites and aplites; (2) Li-poor granitoids and associated pegmatites and aplites; (3) Ca-rich metapelites, metapsammites, and calc-silicate rocks; (4) Ca-poor metapelites, metapsammites, and quartz-tourmaline rocks; (5) metacarbonates; (6) metaultramafics.

**Fig. 9**. a) Total Alkali vs Silica classification diagram (Le Bas et al., 1986); b) $Zr/TiO_2$ vs Nb/Y classification diagram (Winchester and Floyd, 1977).

**Fig. 10**. a) Incompatible element distribution of the studied felsic metavolcanic rocks. Element concentrations are normalized to the Primitive Mantle composition of McDonough and Sun (1995). b) REE distribution of the studied samples. Element concentrations are normalized to the CI chondrite composition of McDonough and Sun (1995).

**Fig. 11**. Cathodoluminescence (SEM-CL) images of zircon crystals from the four selected samples. Selected ages are shown.

**Fig. 12**. Histograms, probability density plot, and concordia diagram of LA-ICP-MS U-Pb zircon ages of the Fornovolasco metarhyolite. The complete U-Pb data set is reported in Supplementary Table S1.

**Fig. 13**. Histograms, probability density plots, and concordia diagrams of LA-ICP-MS U-Pb zircon ages for the four samples of the Fornovolasco metarhyolite.

**Fig. 14**. Available geochronological data for the Paleozoic rocks cropping out in the Northern Apennines (in red). Ellipses report ages of magmatic rocks (red), of Variscan metamorphism (green), and the maximum depositional ages, based on detrital zircon (blue). References: (1) this work; (2) Eberhardt et al. (1962); (3) Lo Pò et al. (2016); (4) Molli et al. (2002); (5) Paoli et al. (2017); (6) Ferrara and Tonarini (1985); (7) Del Moro et al. (1982); (8) Sirevaag et al. (2016); (9) Musumeci et al. (2011).

**Table 1**. Details of the metarhyolite rock samples from Alpi Apuane studied in this work.

| Sample | Locality | UTM-E (m)* | UTM-N (m)* | Elevation (m a.m.s.l.) | Texture | Mineralogy | Tourmaline-bearing spots |
|---|---|---|---|---|---|---|---|
| Fornovolasco Metarhyolite Fm | | | | | | | |
| FVb1 | Fornovolasco/Boscaccio | 609552 | 4875911 | 475 | Gra., unfoliated | Qz, Ab, Wm, Tur | No |
| FVb2 | Fornovolasco/Boscaccio | 609830 | 4876168 | 610 | Gra., slightly foliated | Qz, Kfs, Ab, Wm, Bt, Tur | Common (< 1 |
| FVb7 | Fornovolasco/Boscaccio | 609844 | 4876270 | 570 | Por., unfoliated | Qz, Kfs, Ab, Wm, Bt, Tur, Chl | Rare (< 1 cm |
| FVb9 | Fornovolasco/Boscaccio | 609684 | 4876496 | 575 | Por., slightly foliated | Qz, Kfs, Ab, Wm, Bt, Tur, Chl | Rare (< 5 mm |
| FVb13 | Fornovolasco/Boscaccio | 609853 | 4876280 | 570 | Por., unfoliated | Qz, Kfs, Ab, Wm, Bt, Tur, Chl | Common (< |
| FVc1 | Fornovolasco/Le Casette | 608227 | 4875263 | 600 | Gra., unfoliated | Qz, Ab, Wm, Tur | No |
| FVc2 | Fornovolasco/Le Casette | 608253 | 4875258 | 585 | Gra., unfoliated | Qz, Wm, Tur | No |
| FVc8 | Fornovolasco/Le Casette | 608253 | 4875259 | 585 | Gra., unfoliated | Qz, Ab, Wm, Tur | No |
| FVt3 | Fornovolasco/Trimpello | 609198 | 4876438 | 675 | Por., slightly foliated | Qz, Kfs, Ab, Wm, Tur, Chl | Rare (< 2 cm |
| POL1 | Valdicastello Carducci | 602240 | 4868881 | 290 | Por., slightly foliated | Qz, Ab, Wm, Tur | Common (< |
| MUL1 | Mulina di Stazzema | 604858 | 4870984 | 410 | Por., unfoliated | Qz, Ab Wm, Tur | Rare (< 1 cm |
| SAS1 | Sant'Anna di Stazzema | 602184 | 4869898 | 650 | Por., slightly foliated | Qz, Ab, Wm, Bt, Tur, Chl | Rare (< 2 cm |
| "*Porfiroidi e scisti porfirici*" Fm | | | | | | | |
| PORF1 | Passo del Pitone, Massa | 597922 | 4878168 | 1220 | Por., strongly foliated | Qz, Kfs, Ab, Wm | No |
| PORF2 | Monte dei Ronchi, Seravezza | 600292 | 4878164 | 1275 | Por., strongly foliated | Qz, Wm | No |
| PORF3 | Fociomboli, Stazzema | 602613 | 4877421 | 1230 | Por., strongly foliated | Qz, Kfs, Ab Wm | No |

Abbreviations: Ab, albite; Bt, biotite; Chl, chlorite; Kfs, K-feldspar; Qz, quartz; Tur, tourmaline; Wm, white mica. Gra., granular; Por., porphyritic; a.m.s.l., above mean s

*Coordinate system: WGS84-UTM32N

**Table 2**. Chemical analyses of phyllosilicates occurring in the Fornovolasco metarhyolite.

| | biotite | | chlorite | | | | | | muscovite | | | |
|---|---|---|---|---|---|---|---|---|---|---|---|---|
| | FVb7 (n = 28) | | SAS1 (n = 37) | | FVb7 dark (n = 7) | | FVb7 bright (n = 19) | | FVb9 dark (n = 12) | | | |
| | wt% | e.s.d. | wt% | e.s.d. | wt% | e.s.d. | wt% | e.s.d. | wt% | e.s.d. | | |
| $SiO_2$ | 36.33 | 0.99 | 24.87 | 0.61 | 46.42 | 0.65 | 48.06 | 1.45 | 46.60 | 1.31 | | |
| $TiO_2$ | 2.70 | 0.24 | 0.07 | 0.04 | 0.05 | 0.02 | 0.41 | 0.16 | 0.39 | 0.16 | | |
| $Al_2O_3$ | 13.95 | 0.44 | 20.34 | 0.64 | 34.71 | 1.20 | 24.19 | 0.86 | 30.91 | 1.92 | | |
| $Cr_2O_3$ | 0.04 | 0.03 | - | - | 0.02 | 0.01 | - | - | - | - | | |
| $V_2O_3$ | 0.29 | 0.04 | 0.03 | 0.02 | - | - | - | - | - | - | | |
| MgO | 8.56 | 0.29 | 15.13 | 0.53 | 0.54 | 0.25 | 3.26 | 0.44 | 1.86 | 0.68 | | |
| CaO | - | - | 0.03 | 0.02 | 0.03 | 0.02 | - | - | - | - | | |
| MnO | 0.05 | 0.03 | 0.18 | 0.03 | 0.05 | 0.04 | - | - | - | - | | |
| FeO | 23.26 | 0.49 | 24.49 | 0.63 | 1.24 | 0.60 | 6.02 | 1.45 | 2.16 | 0.42 | | |
| SrO | - | - | - | - | - | - | - | - | - | - | | |
| BaO | - | - | - | - | 0.08 | 0.04 | 0.96 | 0.41 | 0.08 | 0.11 | | |
| $Na_2O$ | 0.03 | 0.02 | - | - | 0.11 | 0.08 | 0.04 | 0.02 | 0.19 | 0.03 | | |
| $K_2O$ | 8.83 | 0.34 | - | - | 10.82 | 0.23 | 10.24 | 0.42 | 10.30 | 0.15 | | |
| F | 0.13 | 0.03 | - | - | - | - | - | - | - | - | | |
| $H_2O_{calc}$* | 3.75 | | 11.05 | | 4.45 | | 4.25 | | 4.36 | | | |
| O = F | -0.05 | | - | | | | | | | | | |
| Total | 97.86 | | 96.16 | | 98.52 | | 97.43 | | 96.85 | | | |
| apfu | normalized to O = 12 | | normalized to O = 18 | | normalized to O = 12 | | normalized to O = 12 | | normalized to O = 12 | | | |
| $Si^{4+}$ | 2.86 | | 2.70 | | 3.13 | | 3.38 | | 3.21 | | | |
| $^{IV}Al^{3+}$ | 1.14 | | 1.30 | | 0.87 | | 0.62 | | 0.79 | | | |
| $Ti^{4+}$ | 0.16 | | 0.01 | | - | | 0.02 | | 0.02 | | | |
| $^{VI}Al^{3+}$ | 0.15 | | 1.30 | | 1.89 | | 1.39 | | 1.72 | | | |
| $Cr^{3+}$ | - | | - | | - | | - | | - | | | |
| $V^{3+}$ | 0.02 | | - | | - | | - | | - | | | |
| $Mg^{2+}$ | 1.00 | | 2.45 | | 0.05 | | 0.34 | | 0.19 | | | |
| $Ca^{2+}$ | - | | - | | - | | - | | - | | | |
| $Mn^{2+}$ | - | | 0.02 | | - | | - | | - | | | |
| $Fe^{2+}$ | 1.53 | | 2.22 | | 0.07 | | 0.35 | | 0.12 | | | |
| $Sr^{2+}$ | - | | - | | - | | - | | - | | | |
| $Ba^{2+}$ | - | | - | | - | | 0.03 | | - | | | |
| $Na^+$ | - | | - | | 0.01 | | 0.01 | | 0.01 | | | |
| $K^+$ | 0.89 | | - | | 0.93 | | 0.92 | | 0.90 | | | |
| $F^-$ | 0.03 | | - | | - | | - | | - | | | |
| $OH^-$ | 1.97 | | 8.00 | | 2.00 | | 2.00 | | 2.00 | | | |

* $H_2O$ (wt%) calculated on the basis of stoichiometry. e.s.d., estimated standard deviation.

**Table 3**. Selected chemical analyses of tourmaline occurring in the Fornovolasco metarhyolite.

| | FVb7 | | | | FVb9 | | | | | |
|---|---|---|---|---|---|---|---|---|---|---|
| | dark brown | blue | blue | dark brown | brown | blue | pale blue | pale blue | brown | tan |
| | wt% | wt% | wt% | wt% | wt% | wt% | wt% | wt% | wt% | wt% |
| $SiO_2$ | 35.41 | 33.62 | 36.23 | 35.51 | 35.89 | 36.05 | 37.99 | 37.38 | 34.29 | 35.93 |
| $TiO_2$ | 4.22 | 0.10 | 0.25 | 4.62 | 0.05 | 0.04 | 0.27 | 0.11 | 1.32 | 1.76 |
| $Al_2O_3$ | 26.09 | 30.94 | 32.24 | 26.20 | 35.51 | 36.53 | 38.10 | 34.78 | 30.43 | 32.16 |
| $V_2O_3$ | 0.54 | - | 0.05 | 0.51 | 0.04 | 0.02 | 0.05 | - | 0.18 | 0.24 |
| MgO | 6.10 | 4.94 | 5.68 | 6.32 | 2.82 | 2.93 | 8.62 | 8.03 | 6.75 | 6.88 |
| CaO | 1.47 | 0.44 | 0.91 | 1.20 | 0.26 | 0.30 | 0.51 | 0.48 | 1.27 | 0.47 |
| MnO | 0.07 | 0.06 | 0.16 | 0.07 | 0.05 | 0.06 | 0.01 | - | 0.02 | - |
| FeO | 11.91 | 9.49 | 9.49 | 11.99 | 11.82 | 10.28 | 0.58 | 1.95 | 8.05 | 6.39 |
| SrO | 0.15 | 0.06 | - | 0.11 | - | 0.04 | 0.23 | 0.04 | - | - |
| $Na_2O$ | 2.19 | 2.00 | 2.13 | 2.36 | 1.85 | 1.75 | 2.24 | 2.54 | 1.92 | 2.16 |
| $K_2O$ | 0.06 | 0.02 | 0.04 | 0.05 | 0.02 | 0.01 | 0.02 | 0.02 | 0.01 | 0.01 |
| Total | 88.21 | 81.67 | 87.18 | 88.94 | 88.31 | 88.01 | 88.62 | 85.33 | 84.24 | 86.00 |
| apfu (normalized to O = 24.5, assuming 3 $BO_3$ groups and 4 OH groups) | | | | | | | | | | |
| *T* cations | | | | | | | | | | |
| $Si^{4+}$ | 5.91 | 5.89 | 5.94 | 5.88 | 5.83 | 5.82 | 5.82 | 5.98 | 5.81 | 5.87 |
| $Al^{3+}$ | 0.09 | 0.11 | 0.06 | 0.12 | 0.17 | 0.18 | 0.18 | 0.02 | 0.19 | 0.13 |
| $\Sigma T$ | 6.00 | 6.00 | 6.00 | 6.00 | 6.00 | 6.00 | 6.00 | 6.00 | 6.00 | 6.00 |
| *Z* cations | | | | | | | | | | |
| $Al^{3+}$ | 5.04 | 6.00 | 6.00 | 4.99 | 6.00 | 6.00 | 6.00 | 6.00 | 5.89 | 6.00 |
| $Mg^{2+}$ | 0.96 | - | - | 1.11 | - | - | - | - | 0.11 | - |
| $\Sigma T$ | 6.00 | 6.00 | 6.00 | 6.00 | 6.00 | 6.00 | 6.00 | 6.00 | 6.00 | 6.00 |
| *Y* cations | | | | | | | | | | |
| $Ti^{4+}$ | 0.53 | 0.01 | 0.03 | 0.57 | 0.01 | - | 0.03 | 0.01 | 0.17 | 0.22 |
| $Al^{3+}$ | - | 0.28 | 0.17 | - | 0.63 | 0.77 | 0.70 | 0.54 | - | 0.07 |
| $V^{3+}$ | 0.07 | - | 0.01 | 0.07 | 0.01 | - | 0.01 | - | 0.02 | 0.03 |
| $Mg^{2+}$ | 0.56 | 1.29 | 1.39 | 0.45 | 0.68 | 0.70 | 1.97 | 1.92 | 1.59 | 1.68 |
| $Mn^{2+}$ | 0.01 | 0.01 | 0.02 | 0.01 | 0.01 | 0.01 | - | - | - | - |
| $Fe^{2+}$ | 1.66 | 1.39 | 1.30 | 1.66 | 1.60 | 1.39 | 0.07 | 0.26 | 1.14 | 0.87 |
| $\Sigma Y$ | 2.83 | 2.98 | 2.92 | 2.76 | 2.94 | 2.87 | 2.71 | 2.73 | 2.92 | 2.87 |
| *X* cations | | | | | | | | | | |
| $Ca^{2+}$ | 0.26 | 0.08 | 0.16 | 0.21 | 0.05 | 0.05 | 0.08 | 0.08 | 0.23 | 0.08 |
| $Sr^{2+}$ | 0.01 | 0.01 | - | 0.01 | - | - | 0.02 | - | - | - |
| $Na^+$ | 0.71 | 0.68 | 0.68 | 0.76 | 0.58 | 0.55 | 0.67 | 0.79 | 0.63 | 0.68 |
| $K^+$ | 0.01 | - | 0.01 | 0.01 | - | - | - | - | - | - |
| $\Sigma X$ | 0.99 | 0.77 | 0.85 | 0.99 | 0.63 | 0.60 | 0.77 | 0.87 | 0.86 | 0.76 |

Note: for every analysis, the natural light color in thin section is given.

**Table 4**. Major- and trace-element analyses of metarhyolite rock samples from Alpi Apuane studied in this work.

|  | *Fornovolasco Metarhyolite Fm* | | | | | | | *"Porfiroidi e scisti porfirici" Fm* | | |
|---|---|---|---|---|---|---|---|---|---|---|
| Sample | FVb1 | FVb2 | FVb13 | FVc2 | FVc8 | SAS1 | MUL1 | PORF1 | PORF2 | PORF3 |
| *Major elements (wt%)* | | | | | | | | | | |
| $SiO_2$ | 73.18 | 74.73 | 69.19 | 73.68 | 74.34 | 69.66 | 75.23 | 67.39 | 76.42 | 70.28 |
| $TiO_2$ | 0.24 | 0.23 | 0.44 | 0.26 | 0.26 | 0.43 | 0.39 | 0.78 | 0.38 | 0.43 |
| $Al_2O_3$ | 13.60 | 14.07 | 15.14 | 14.54 | 14.52 | 14.98 | 14.41 | 15.32 | 13.39 | 14.19 |
| $Fe_2O_{3\,tot}$ | 2.89 | 1.74 | 3.30 | 1.80 | 2.02 | 2.94 | 1.20 | 5.52 | 1.60 | 3.24 |
| MnO | 0.01 | 0.03 | 0.10 | 0.02 | 0.07 | 0.06 | 0.01 | 0.06 | 0.01 | 0.06 |
| MgO | 0.93 | 0.72 | 1.46 | 1.31 | 0.67 | 1.71 | 0.50 | 1.47 | 0.69 | 0.65 |
| CaO | 0.17 | 0.20 | 0.73 | 0.32 | 0.19 | 1.11 | 0.17 | 0.23 | 0.08 | 0.82 |
| $Na_2O$ | 0.29 | 3.92 | 1.81 | 0.27 | 1.14 | 3.84 | 2.85 | 2.98 | 0.14 | 2.59 |
| $K_2O$ | 3.30 | 2.38 | 4.54 | 4.57 | 3.74 | 2.68 | 3.03 | 3.38 | 4.42 | 4.95 |
| $P_2O_5$ | 0.13 | 0.14 | 0.23 | 0.13 | 0.14 | 0.16 | 0.12 | 0.23 | 0.13 | 0.19 |
| LOI | 2.85 | 1.45 | 2.33 | 2.49 | 1.75 | 2.48 | 1.86 | 2.64 | 2.18 | 2.18 |
| Total | 97.59 | 99.61 | 99.27 | 99.39 | 98.84 | 100.05 | 99.77 | 100.00 | 99.44 | 99.58 |
| *Trace elements (ppm)* | | | | | | | | | | |
| Be | 4 | 3 | 4 | 4 | 4 | 3 | 2 | 3 | 3 | 2 |
| Sc | 6 | 4 | 7 | 4 | 4 | 8 | 6 | 13 | 7 | 7 |
| V | 23 | 21 | 47 | 28 | 27 | 49 | 43 | 74 | 26 | 34 |
| Cr | < 20 | < 20 | 50 | < 20 | 30 | 30 | 30 | 50 | < 20 | 30 |
| Co | < 1 | 2 | 11 | < 1 | 4 | 5 | 2 | 12 | 1 | 8 |
| Ni | < 20 | < 20 | 30 | < 20 | < 20 | 20 | < 20 | 20 | < 20 | < 20 |
| Cu | 260 | < 10 | 30 | < 10 | < 10 | < 10 | 40 | 20 | < 10 | 20 |
| Ga | 17 | 14 | 17 | 18 | 18 | 16 | 15 | 20 | 17 | 18 |
| Ge | 2.3 | 2.2 | 2.4 | 2.3 | 1.9 | 1.9 | 1.8 | 1.1 | 1.2 | 1.1 |
| As | > 2000 | 326 | 1120 | 75 | 58 | < 5 | 191 | 10 | 24 | 12 |
| Rb | 116 | 99 | 214 | 176 | 124 | 86 | 110 | 112 | 138 | 111 |
| Sr | 17 | 99 | 69 | 15 | 18 | 106 | 48 | 46 | 18 | 35 |
| Y | 10.9 | 8.4 | 18.2 | 12.3 | 11.5 | 16.6 | 16.4 | 40.9 | 36.6 | 35 |
| Zr | 91 | 76 | 130 | 108 | 93 | 146 | 133 | 321 | 256 | 208 |
| Nb | 7.1 | 7.0 | 7.5 | 8.0 | 7.1 | 7.8 | 6.1 | 14.9 | 11.1 | 11.6 |
| Ag | 1.4 | < 0.5 | 0.6 | 0.6 | < 0.5 | 0.6 | 1.0 | 1.3 | 1.0 | 0.9 |
| Sn | 24 | 10 | 10 | 17 | 17 | 3 | 5 | 3 | 3 | 4 |
| Sb | > 200 | 5.4 | 21.1 | 6.7 | 3.9 | 1.3 | 6.9 | 1.6 | 5.7 | 1.5 |
| Cs | 5.7 | 8.6 | 21.9 | 5.8 | 3.9 | 3.3 | 6.1 | 5.1 | 8.4 | 6.0 |
| Ba | 377 | 933 | 654 | 155 | 139 | 400 | 384 | 684 | 807 | 935 |
| Hf | 2.4 | 2.1 | 3.3 | 2.8 | 2.6 | 3.7 | 3.5 | 8.1 | 6.1 | 5.4 |
| Ta | 1.02 | 1.46 | 0.98 | 1.14 | 1.13 | 1.03 | 0.94 | 1.03 | 0.92 | 0.95 |
| W | 2.0 | 1.7 | 6.5 | 2.0 | 1.4 | 2.1 | 5.9 | 1.7 | 1.7 | 1.6 |
| Tl | 2.68 | 3.84 | 10.4 | 4.19 | 3.58 | 1.21 | 3.57 | 1.09 | 1.02 | 0.75 |
| Pb | 1060 | 48 | 256 | 22 | 13 | 6 | 157 | 21 | 27 | 12 |
| Bi | 58.1 | 3.3 | 50.8 | 38.9 | 1.6 | < 0.1 | 2.1 | 0.8 | 0.3 | 0.1 |
| Th | 9.20 | 5.52 | 9.58 | 10.1 | 9.36 | 10.2 | 9.19 | 19.3 | 18.3 | 15.9 |
| U | 3.18 | 2.69 | 4.44 | 1.82 | 1.72 | 2.05 | 2.11 | 3.03 | 2.94 | 4.13 |
| La | 26.9 | 16.5 | 24.5 | 37.7 | 24.7 | 21.1 | 26.1 | 55.5 | 62 | 47.1 |
| Ce | 54.4 | 34.8 | 50.8 | 76.5 | 48.2 | 42.3 | 49.0 | 118 | 125 | 95.4 |
| Pr | 5.93 | 3.91 | 5.70 | 8.51 | 5.56 | 4.9 | 5.98 | 13.3 | 13.9 | 10.9 |
| Nd | 21.1 | 14.0 | 20.6 | 31.0 | 20.6 | 17.9 | 22.5 | 50.4 | 51.1 | 40.7 |
| Sm | 4.48 | 2.97 | 4.56 | 6.47 | 4.34 | 3.97 | 4.58 | 10.4 | 9.78 | 8.50 |
| Eu | 0.64 | 0.692 | 0.966 | 0.949 | 0.621 | 1.04 | 0.741 | 1.38 | 1.23 | 1.23 |
| Gd | 2.69 | 2.28 | 3.93 | 4.48 | 3.05 | 3.48 | 3.79 | 9.01 | 7.84 | 7.55 |
| Tb | 0.37 | 0.33 | 0.56 | 0.57 | 0.43 | 0.55 | 0.60 | 1.35 | 1.11 | 1.18 |
| Dy | 1.88 | 1.73 | 3.12 | 2.58 | 2.12 | 3.12 | 3.25 | 7.58 | 6.30 | 6.64 |
| Ho | 0.34 | 0.30 | 0.56 | 0.41 | 0.37 | 0.58 | 0.6 | 1.44 | 1.2 | 1.23 |
| Er | 0.99 | 0.81 | 1.60 | 1.06 | 1.05 | 1.55 | 1.75 | 4.03 | 3.45 | 3.39 |
| Tm | 0.155 | 0.119 | 0.214 | 0.134 | 0.147 | 0.227 | 0.235 | 0.585 | 0.504 | 0.455 |
| Yb | 1.04 | 0.77 | 1.38 | 0.83 | 0.99 | 1.60 | 1.57 | 3.89 | 3.32 | 2.93 |
| Lu | 0.166 | 0.123 | 0.228 | 0.131 | 0.152 | 0.234 | 0.248 | 0.560 | 0.496 | 0.475 |

**Table 5**. Summary of LA-ICP-MS U-Pb zircon datings of the Fornovolasco metarhyolite.

| Sample | Zircon crystals imaged by SEM-CL | Zircon crystals selected for U-Pb dating | LA spots | Number of U-Pb concordant ages* | | Age (Ma) | | |
|---|---|---|---|---|---|---|---|---|
| | | | | | | weighted average (±2σ) | min (±2σ) | max (±2σ) |
| FVb13 | 116 | 82 | 122 | rims | 36 | 271.0 (±3.5) | 241.8 (±4.0) | 300.8 (±4.7) |
| | | | | cores | 10 | 276.8 (±8.8) | 260.3 (±7.2) | 298.1 (±4.5) |
| FVc1 | 80 | 61 | 85 | rims | 40 | 279.7 (±3.4) | 256.0 (±6.2) | 296.4 (±3.5) |
| | | | | cores | 7 | 282.6 (±6.6) | 270.0 (±5.6) | 293.7 (±5.5) |
| POL1 | 107 | 55 | 73 | rims | 24 | 276.7 (±8.5) | 226.5 (±6.3) | 306.1 (±6.3) |
| | | | | cores | 0 | - | - | - |
| SAS1 | 114 | 80 | 106 | rims | 44 | 291.8 (±3.2) | 258.6 (±6.6) | 310.6 (±5.4) |
| | | | | cores | 8 | 290.2 (±6.2) | 280.7 (±4.8) | 300.9 (±5.4) |
| | | | | All | 169 | 279.9 (±2.0) | | |

*The cutoff limit was set to 3% normal discordance {[($^{206}$Pb/$^{238}$U age)/($^{207}$Pb/$^{235}$U age) – 1]*100}.

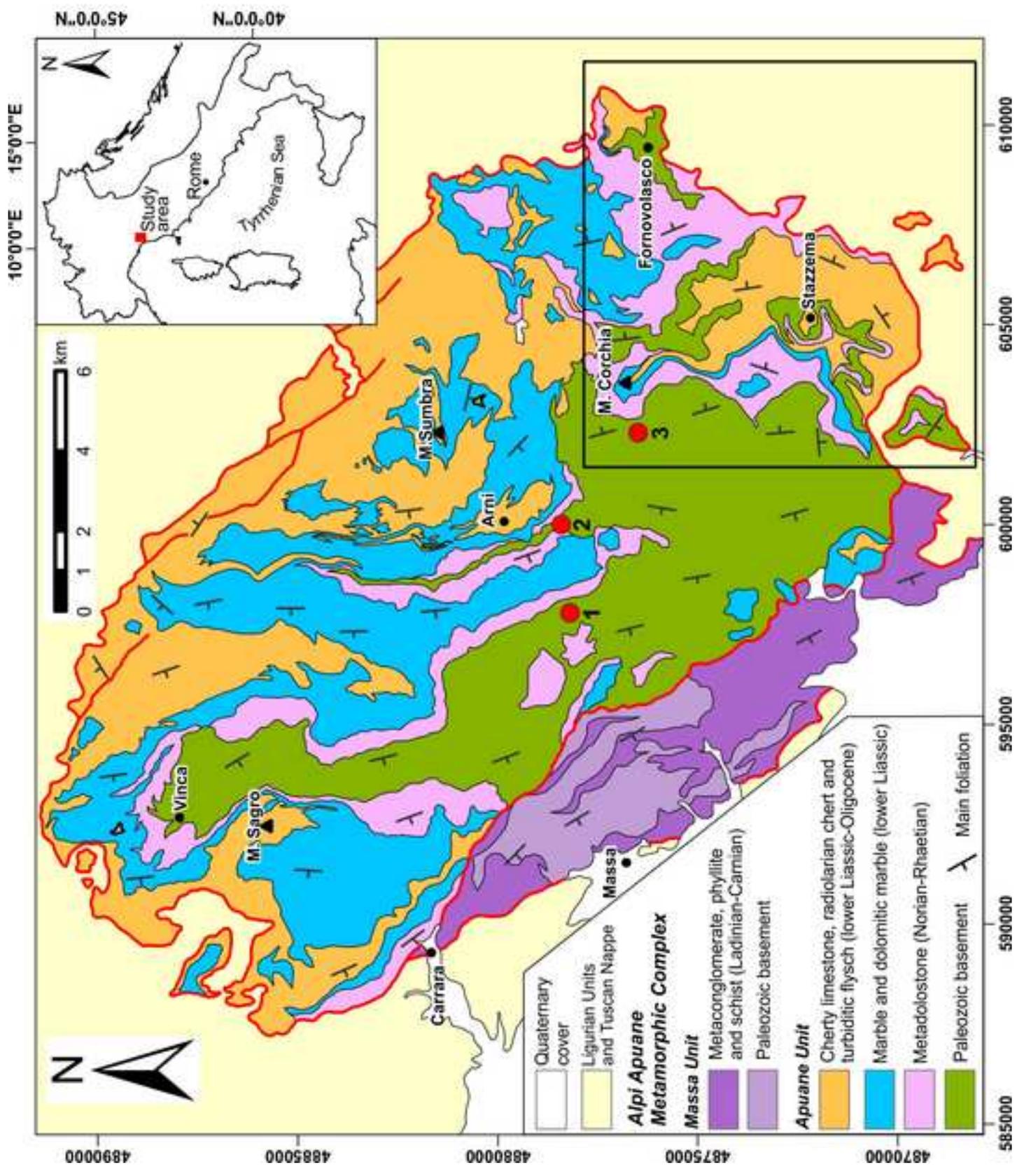

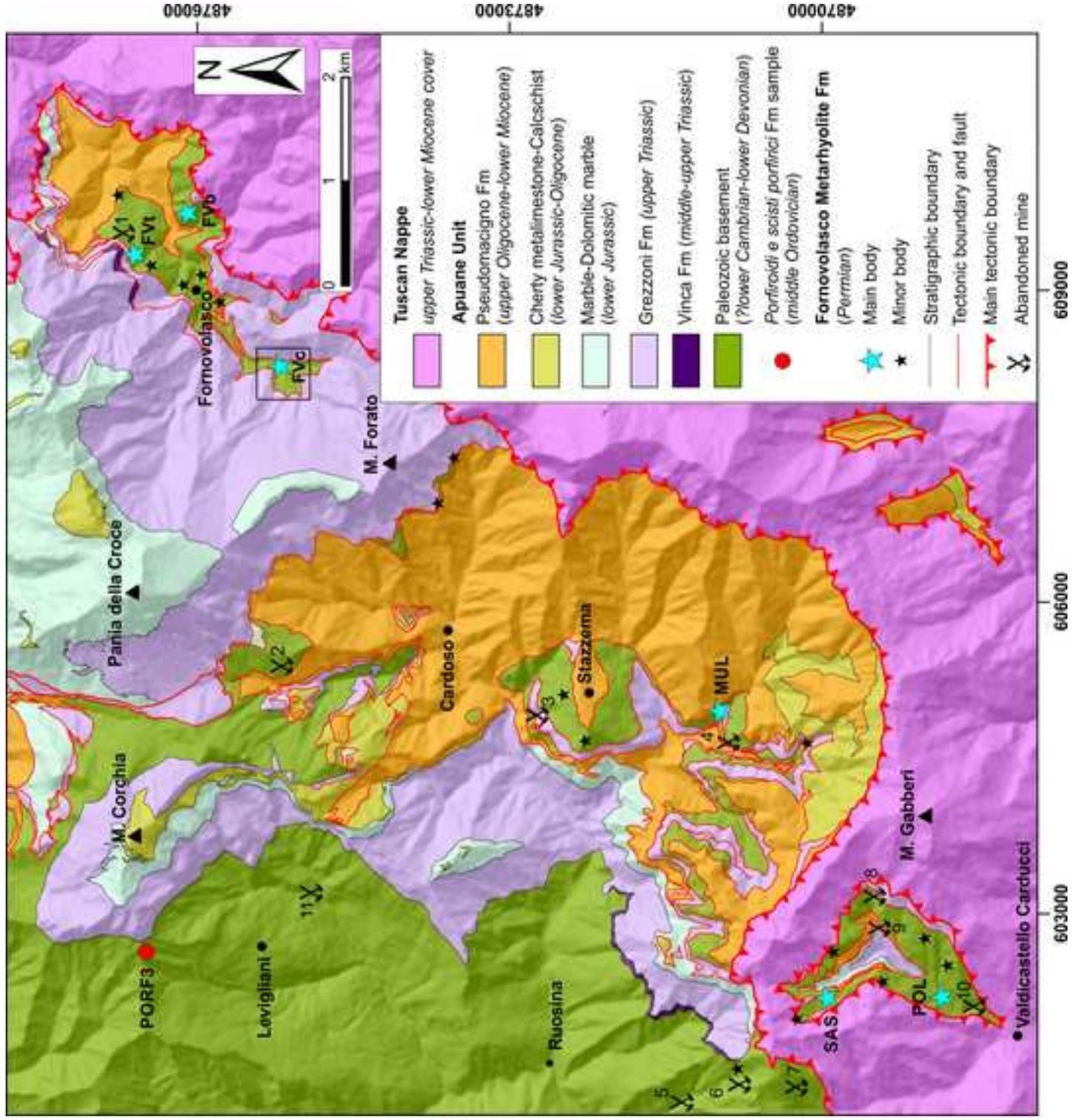

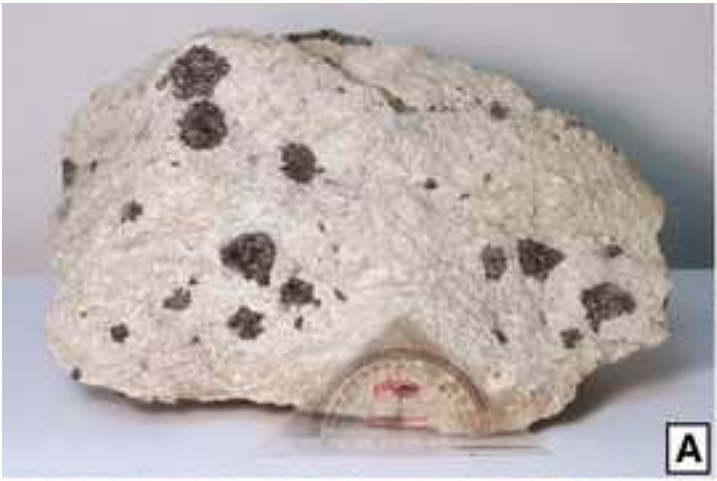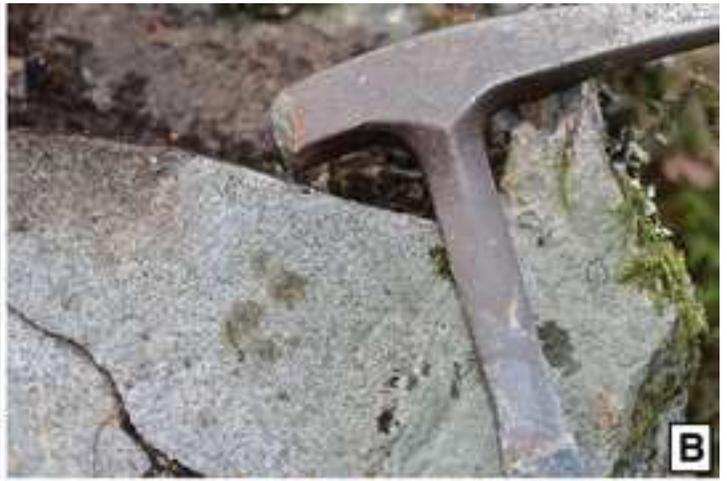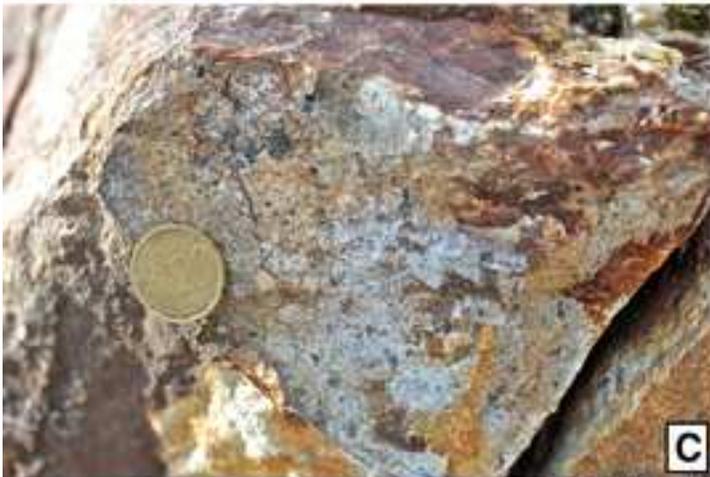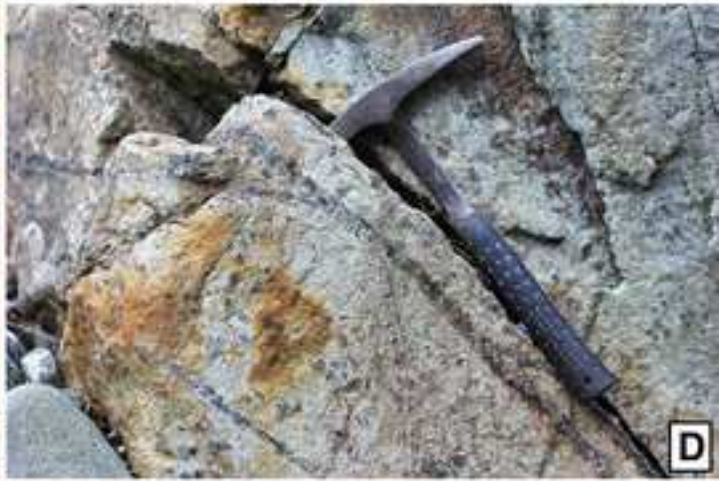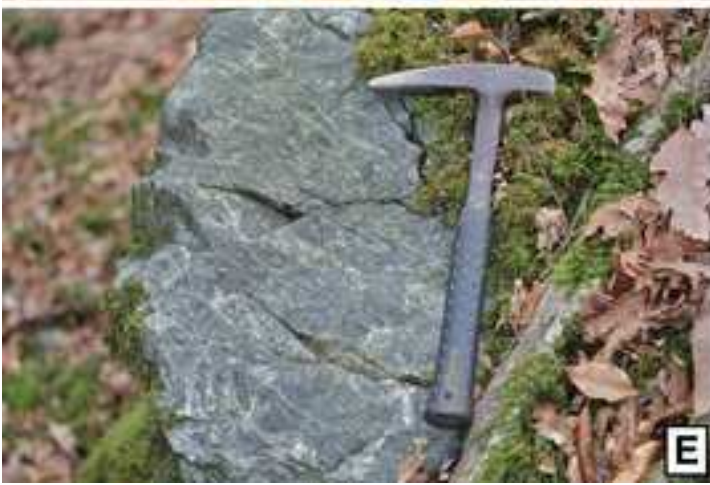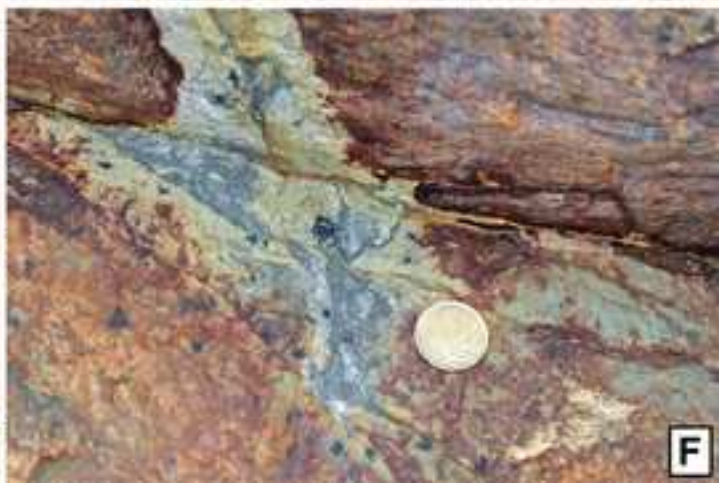

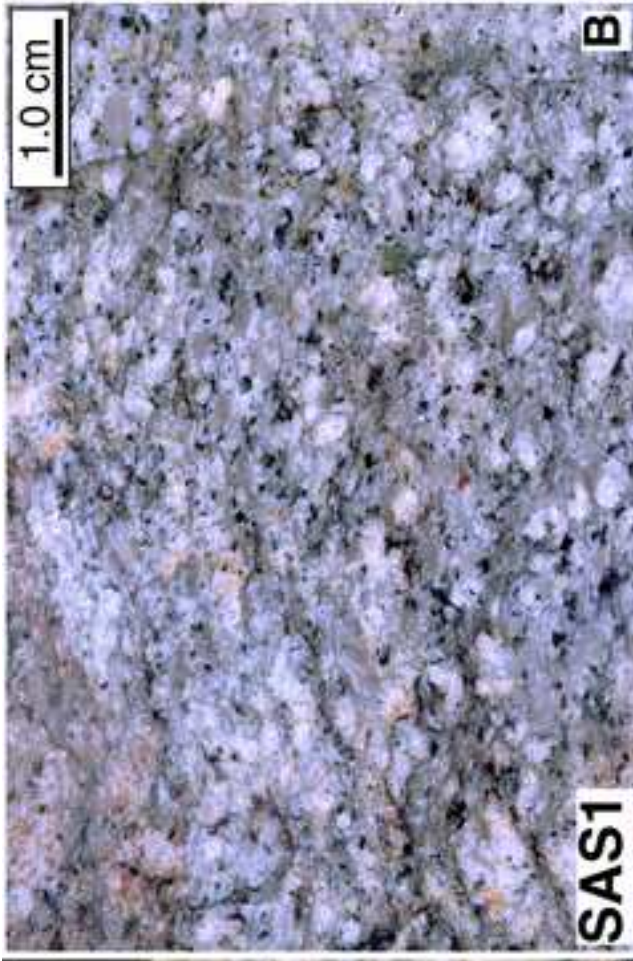
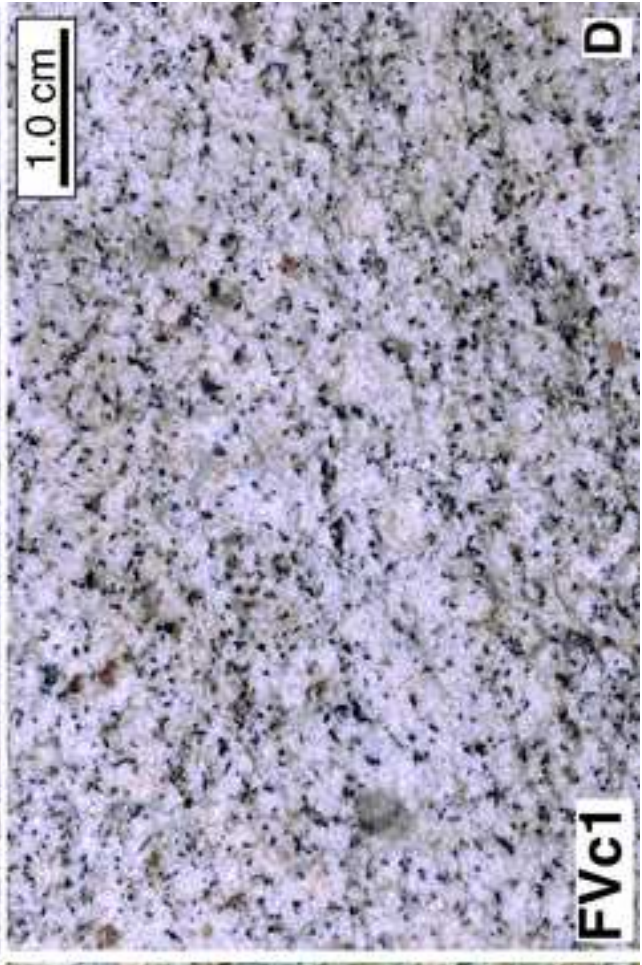
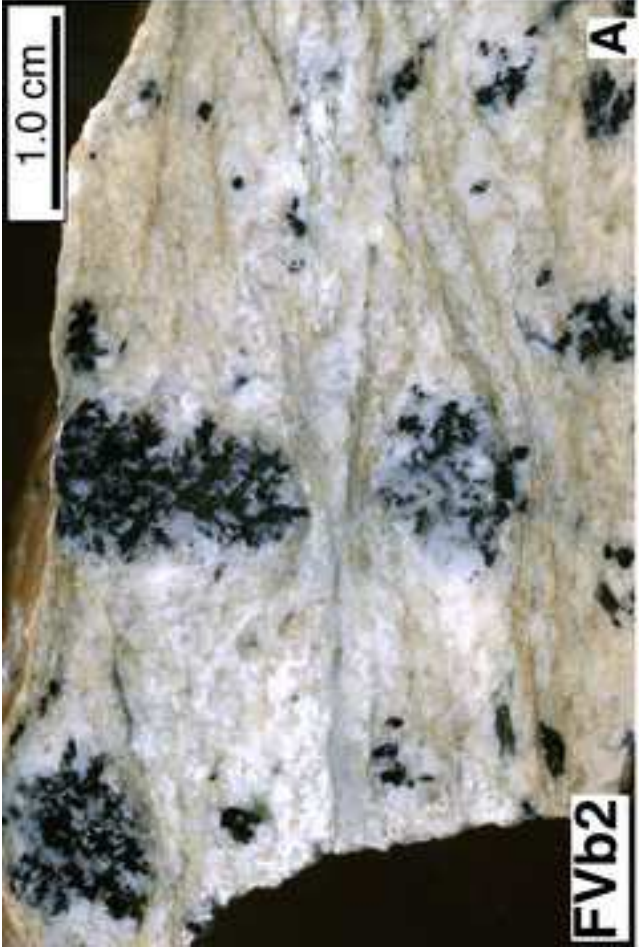
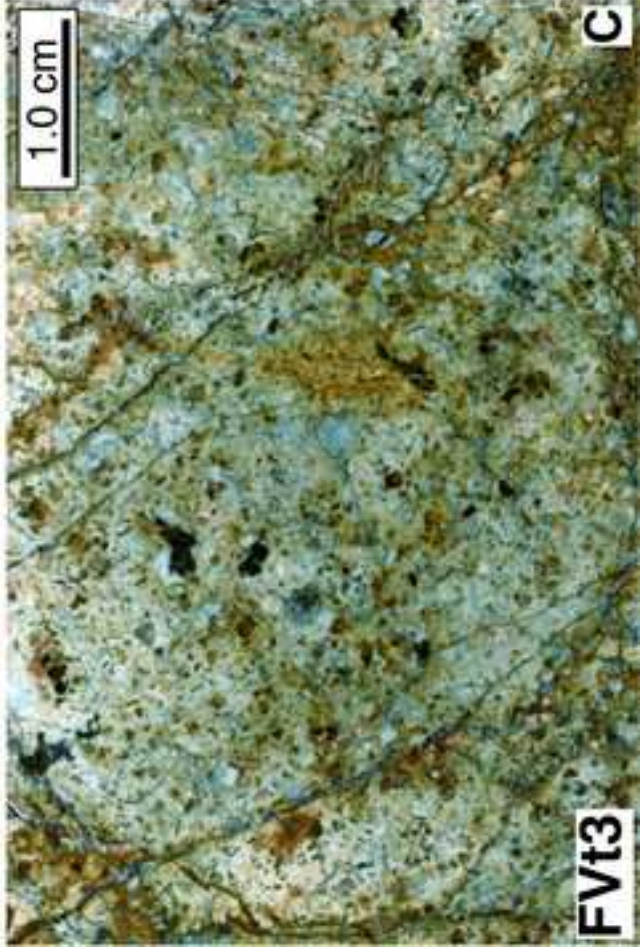

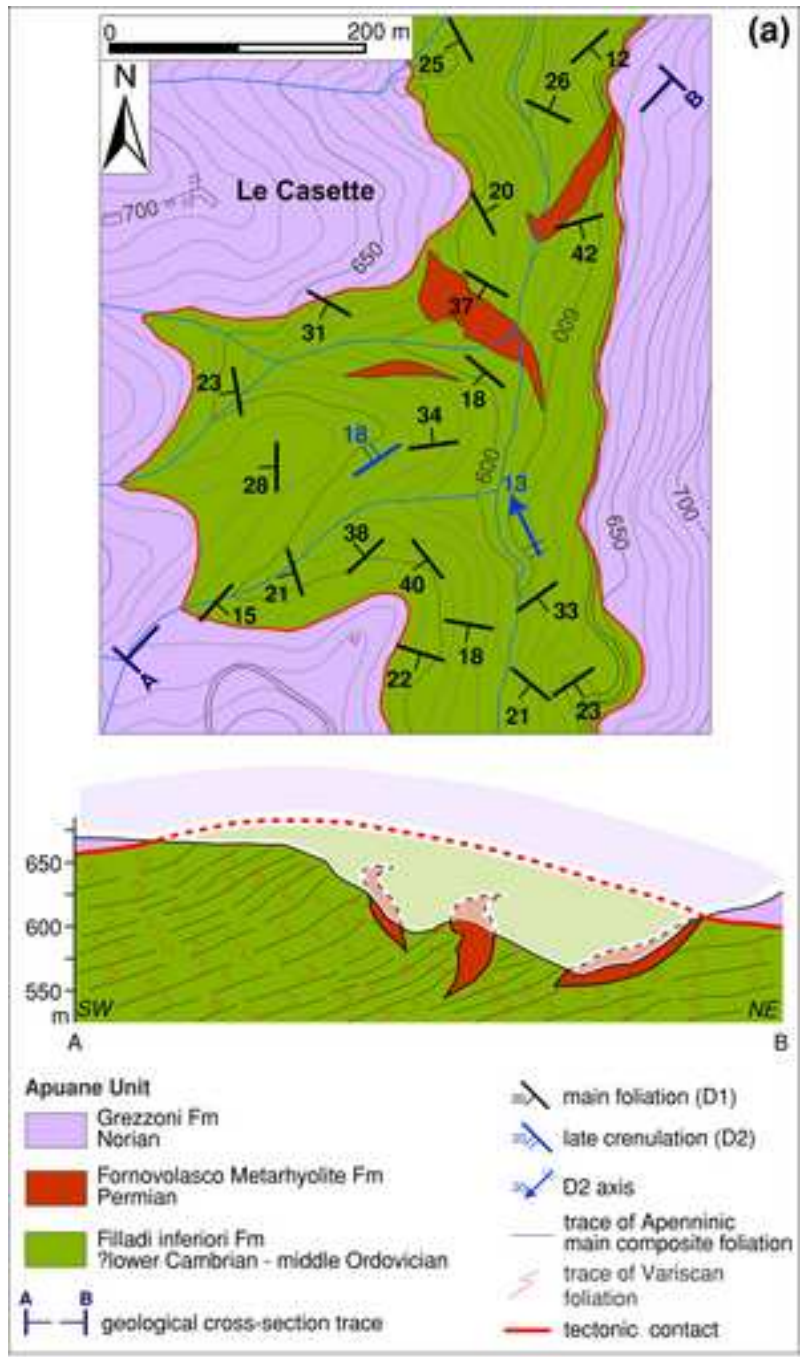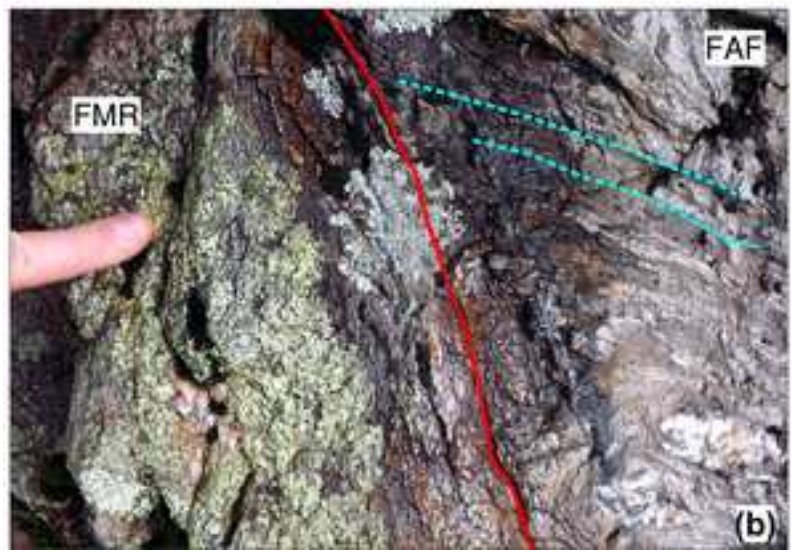

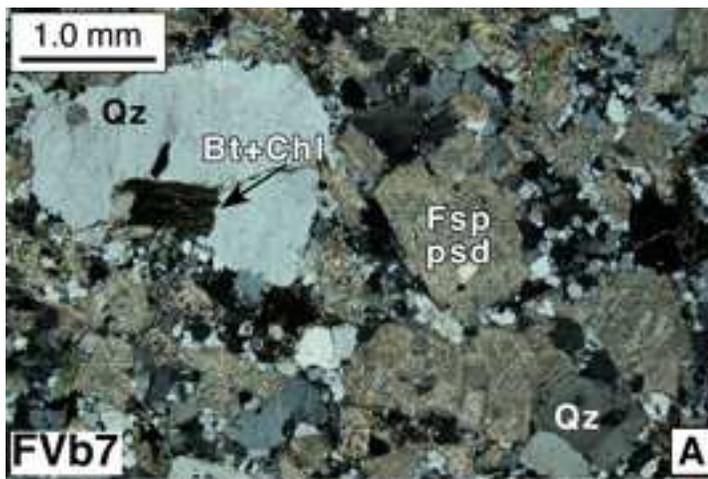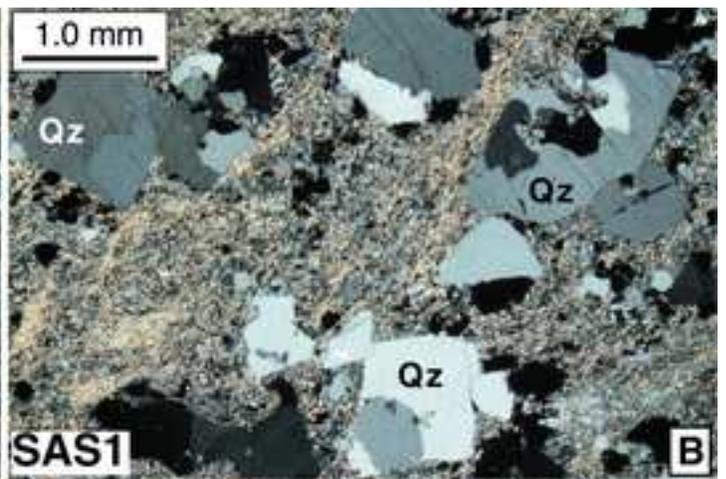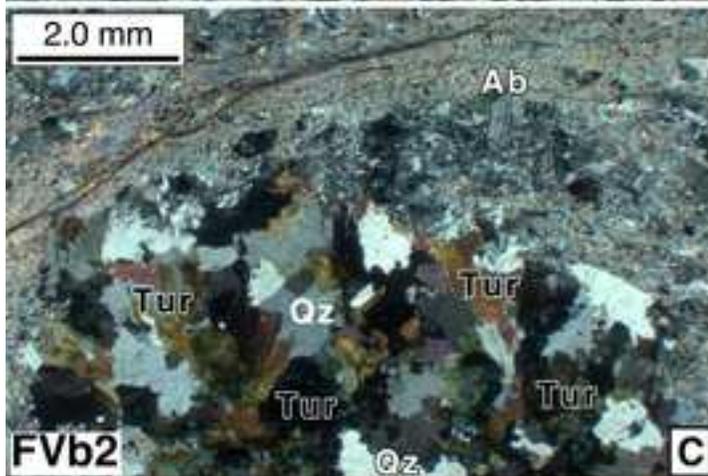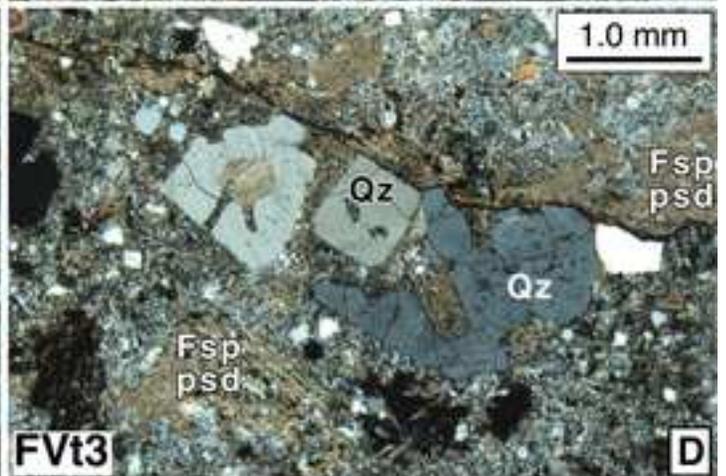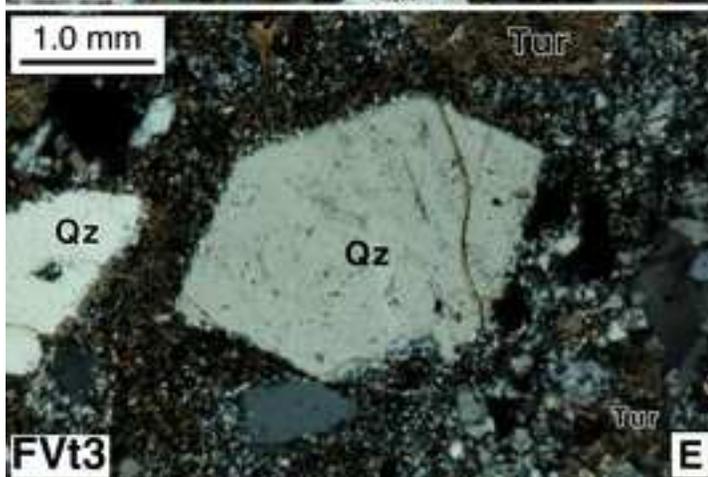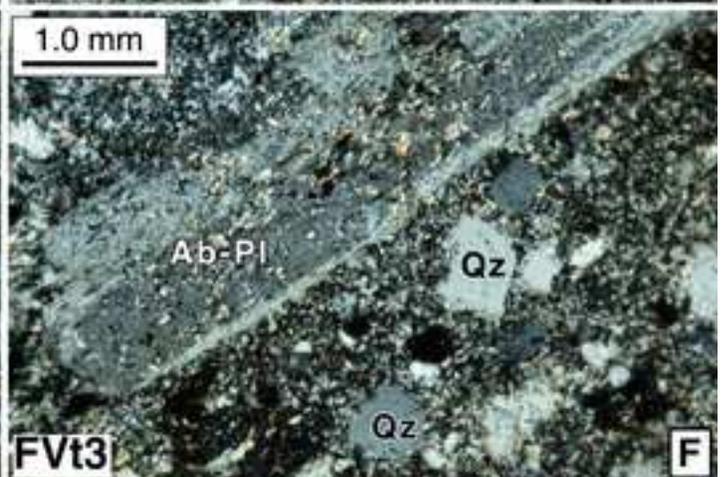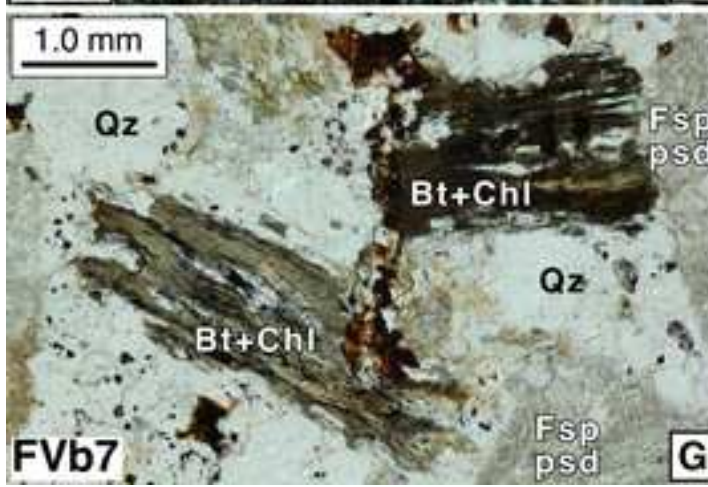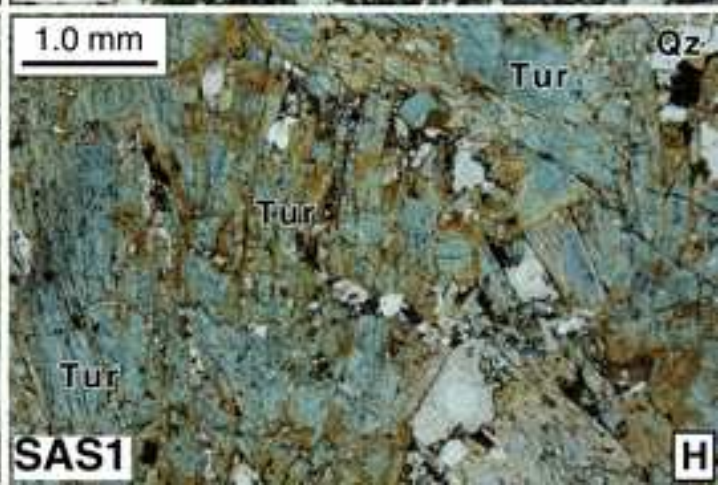

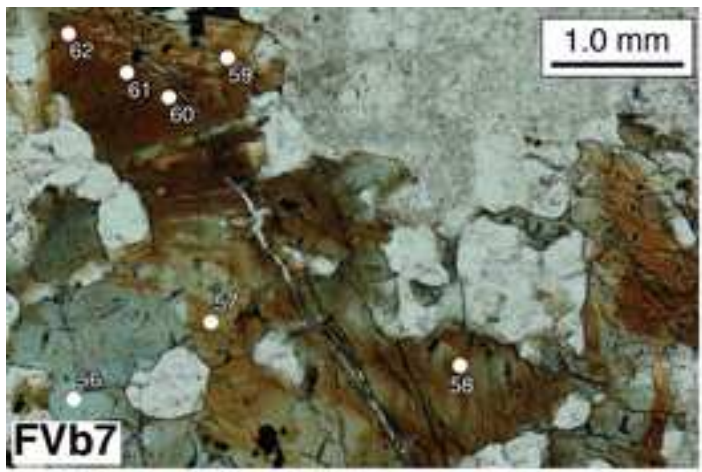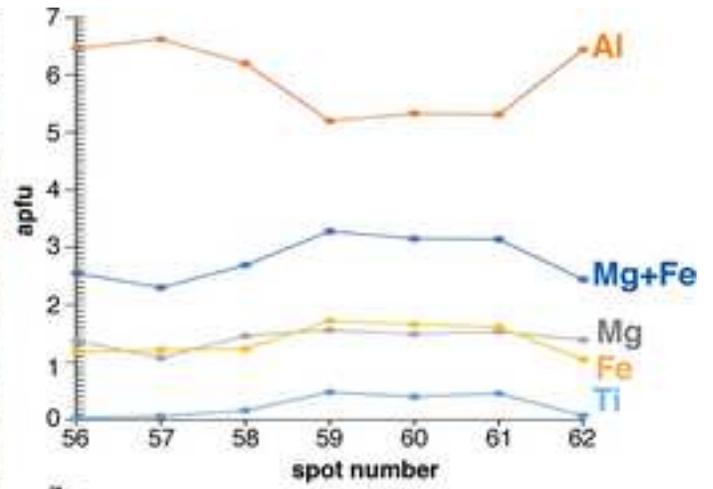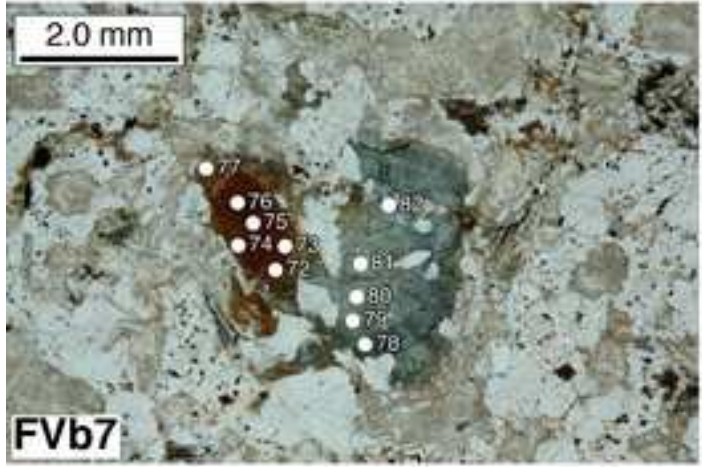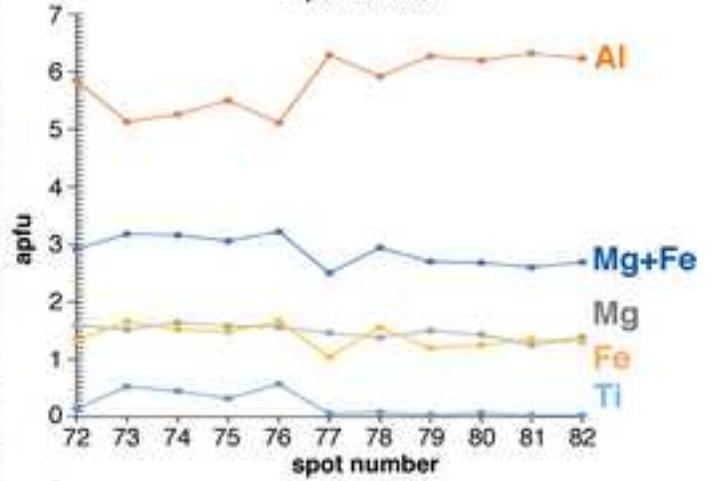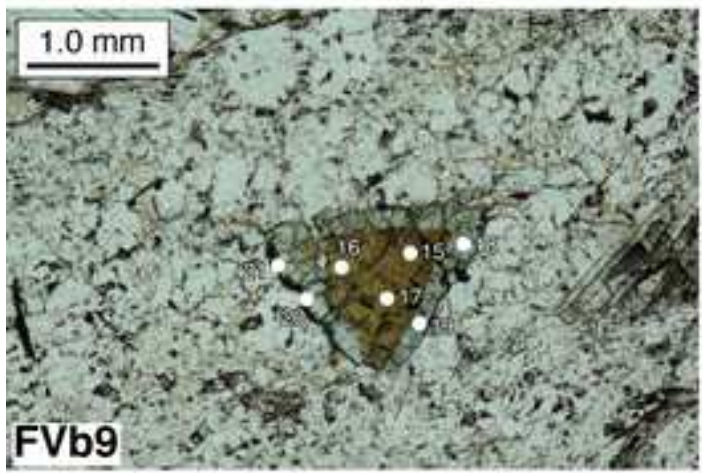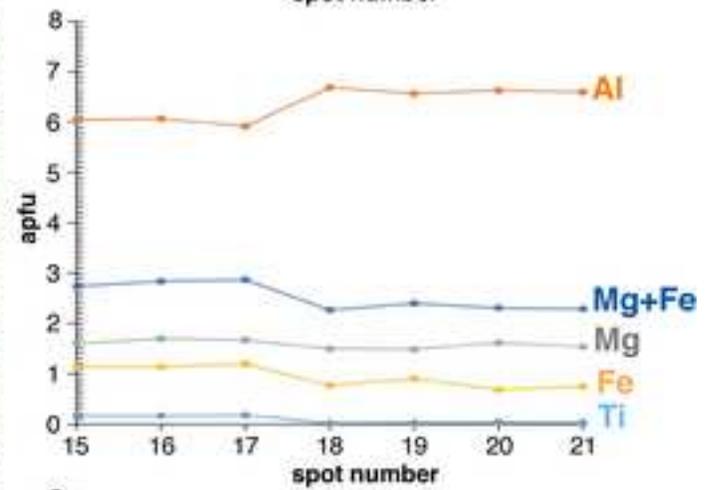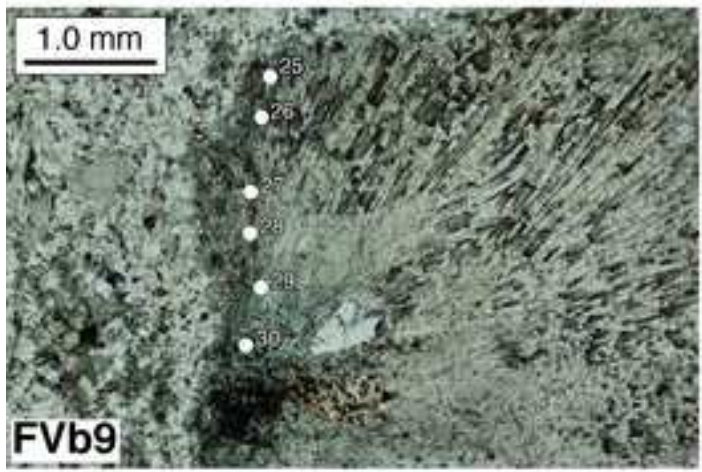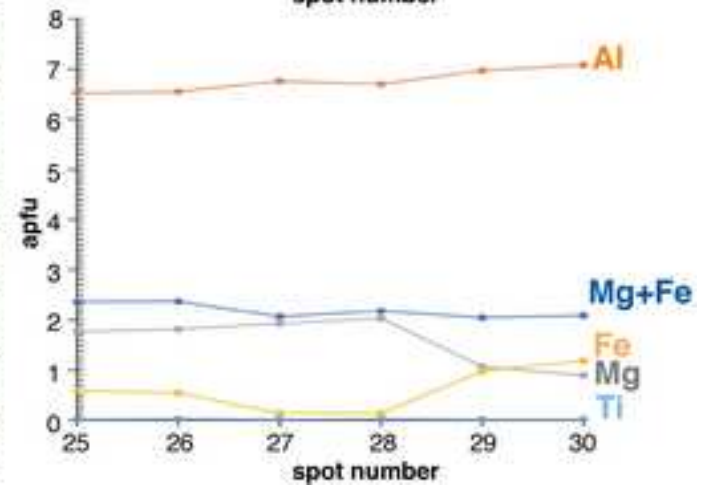

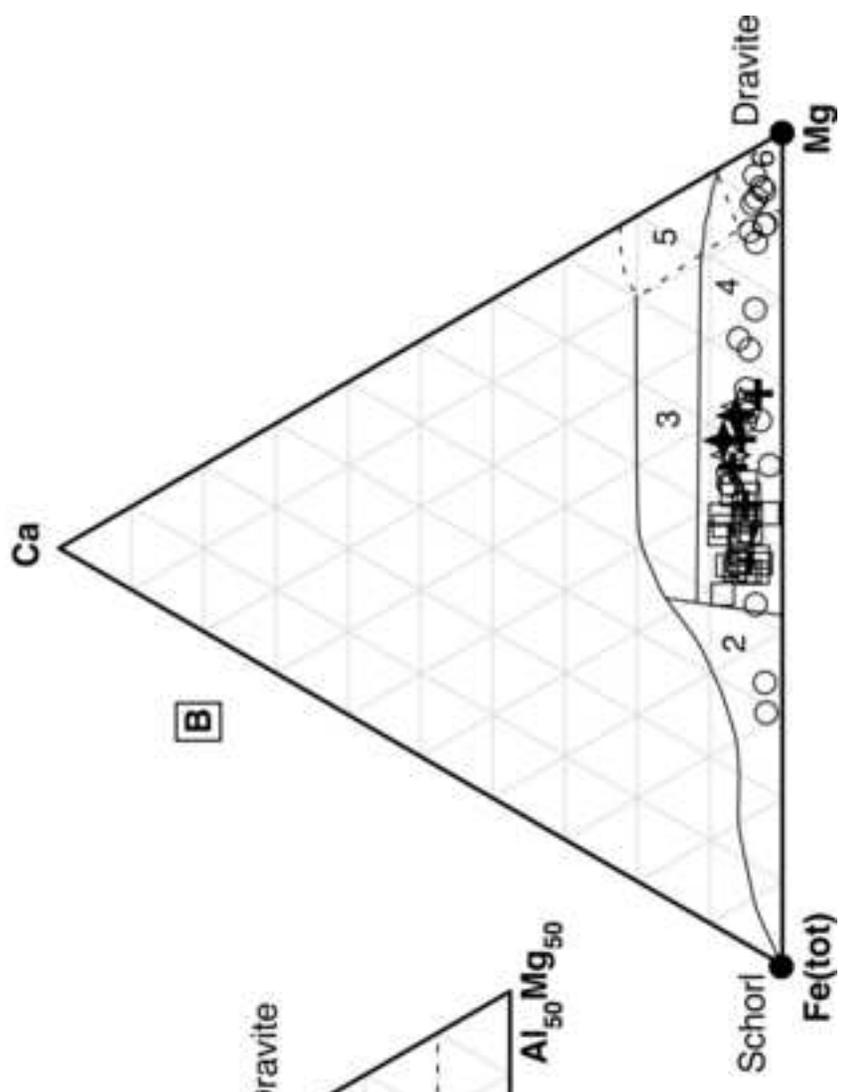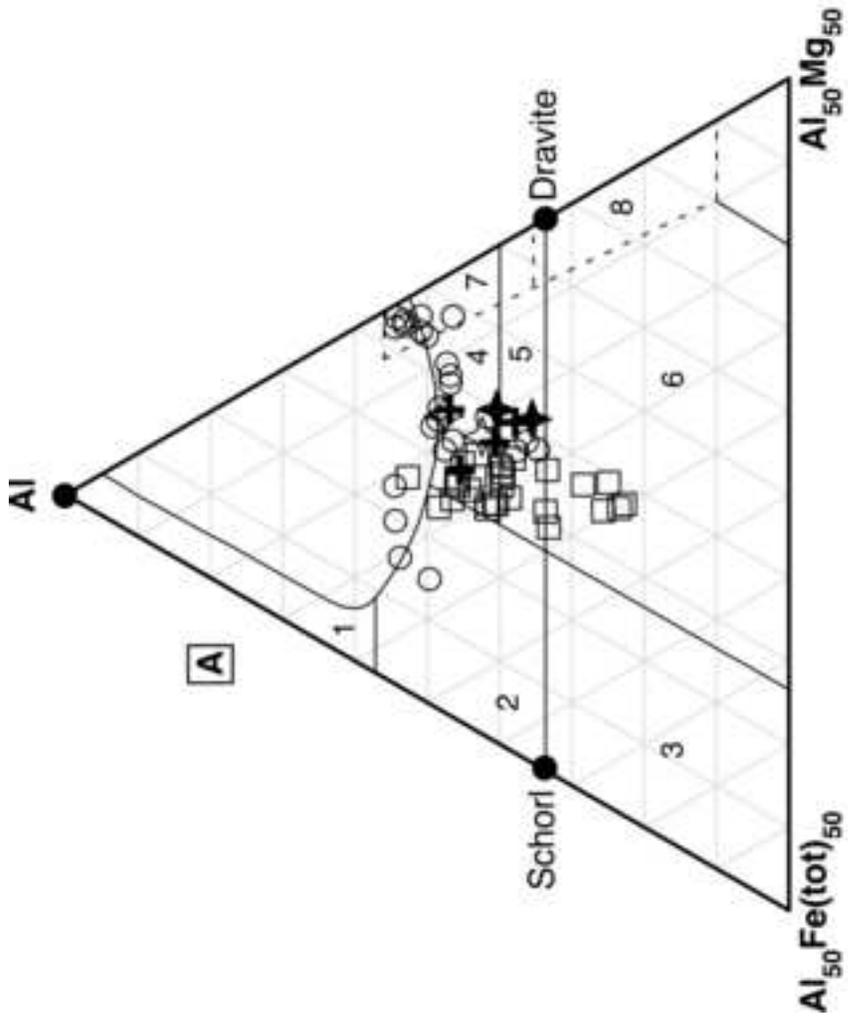

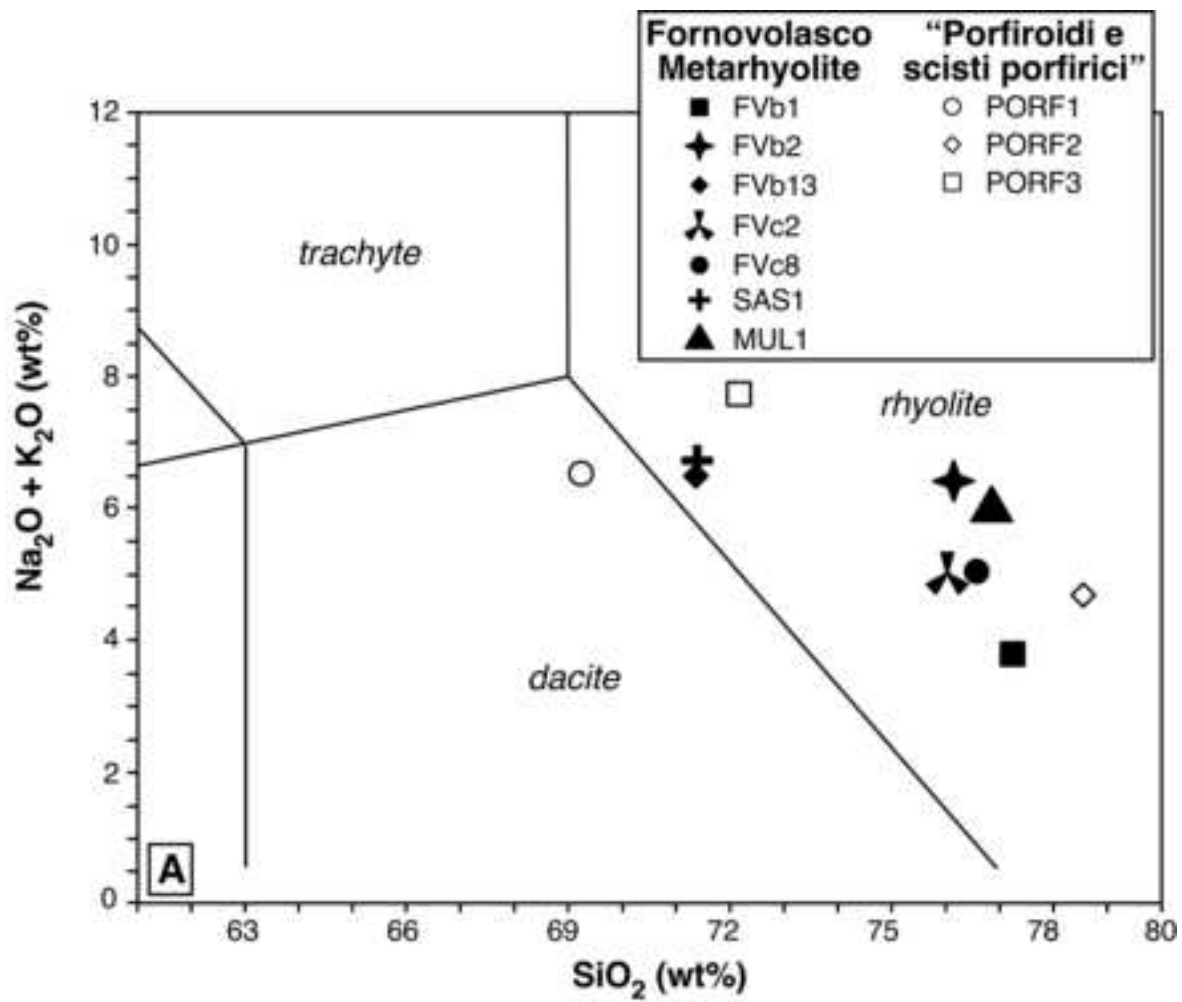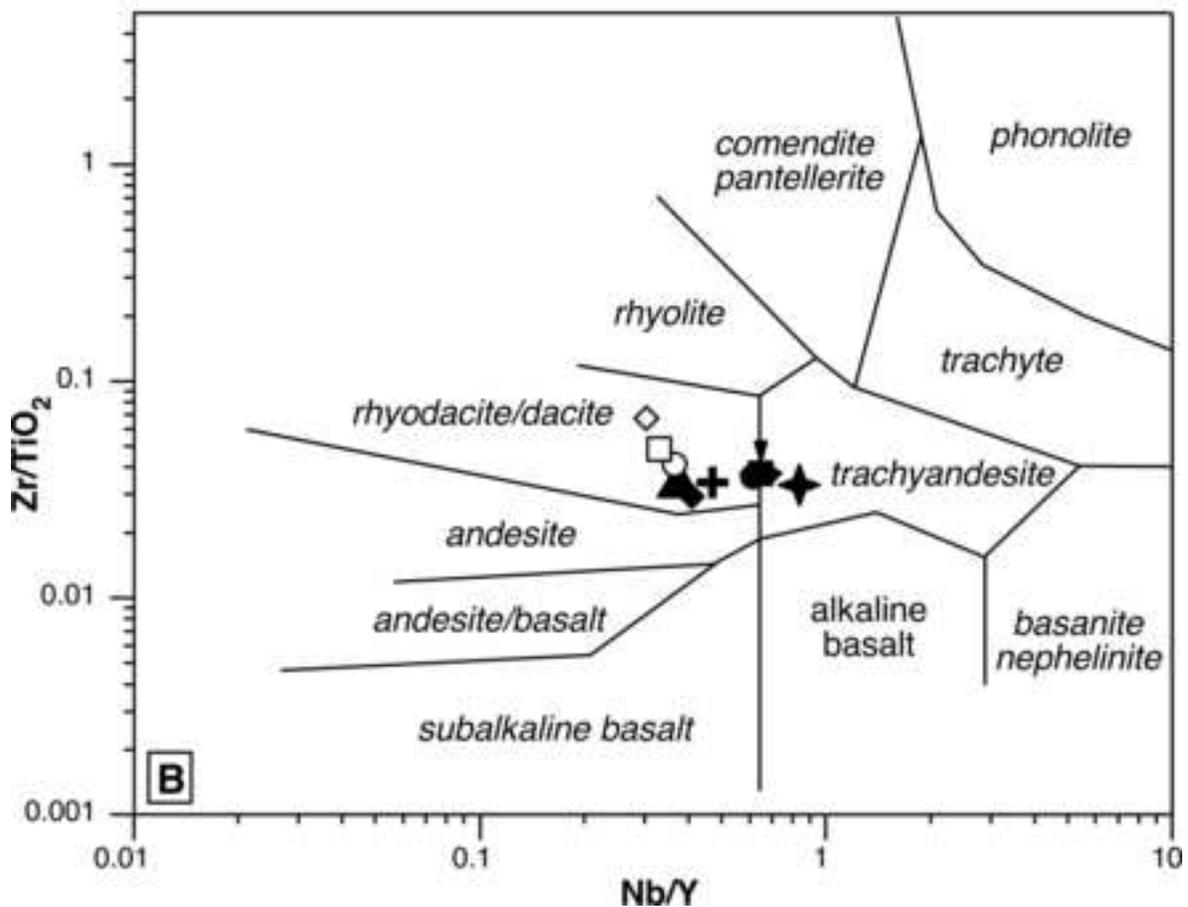

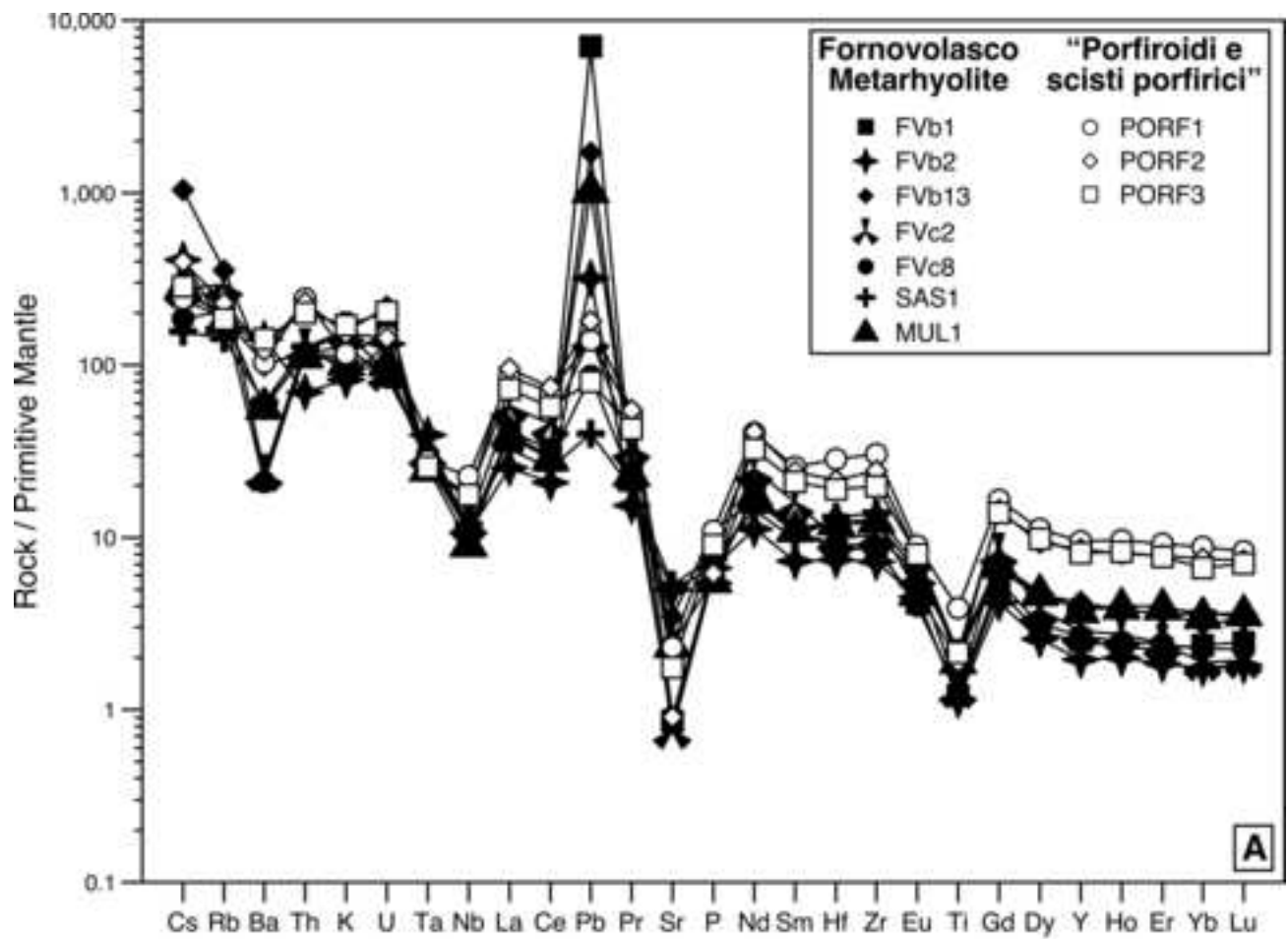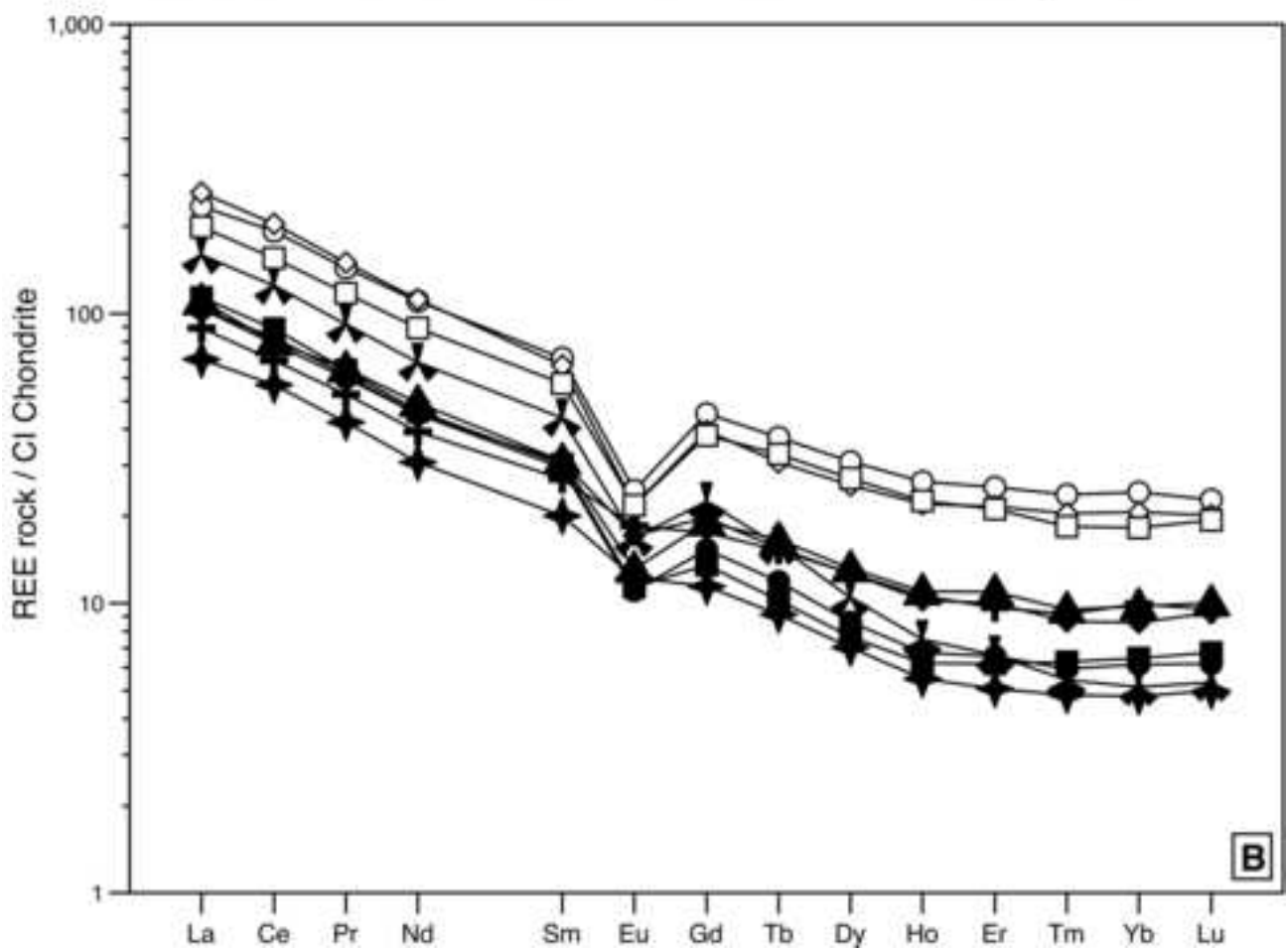

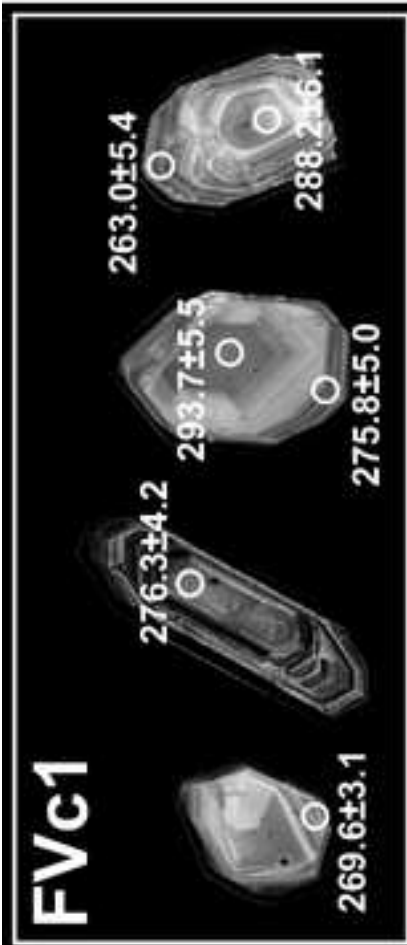
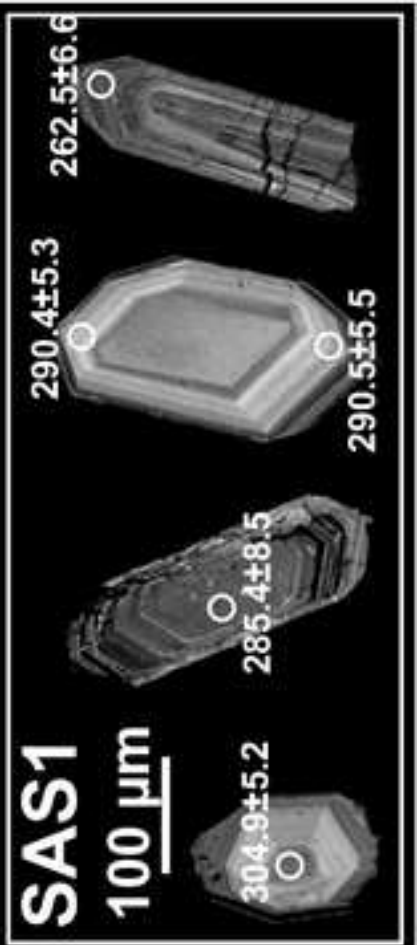
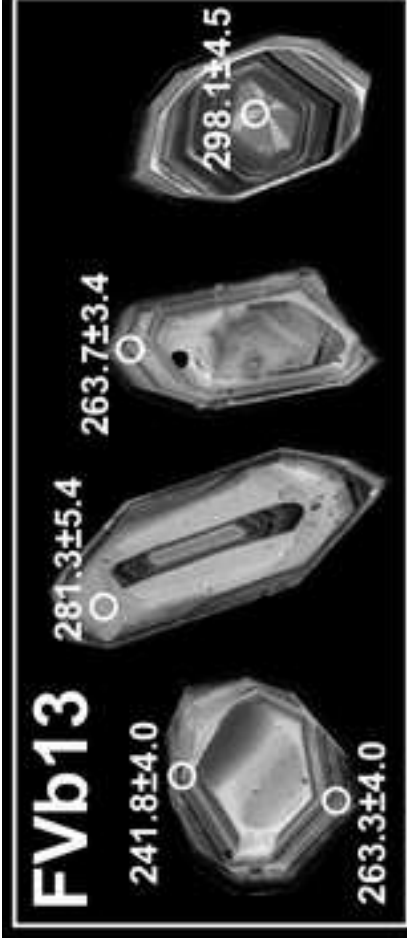
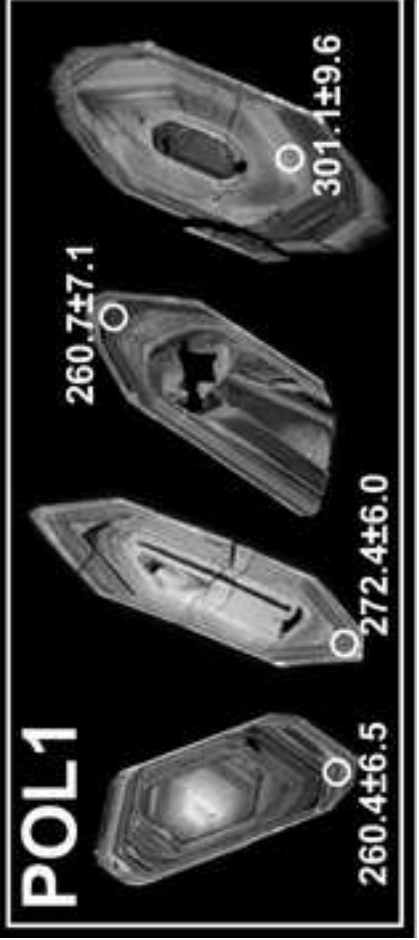

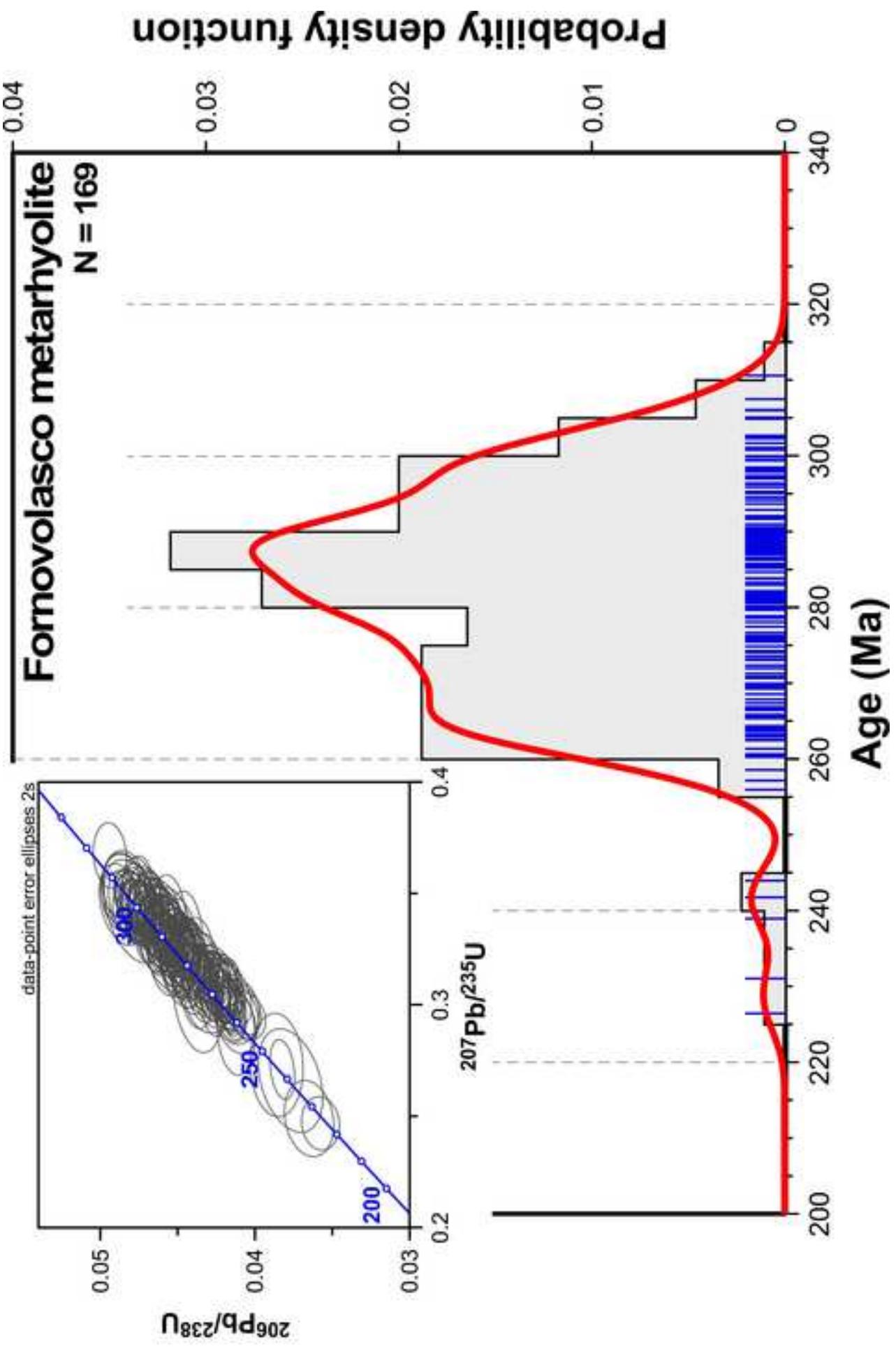

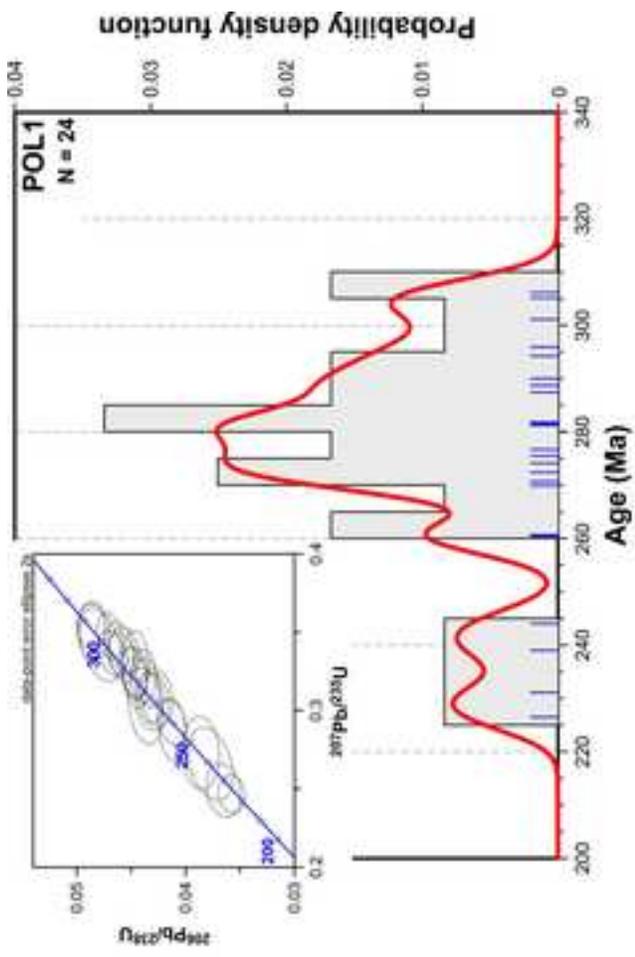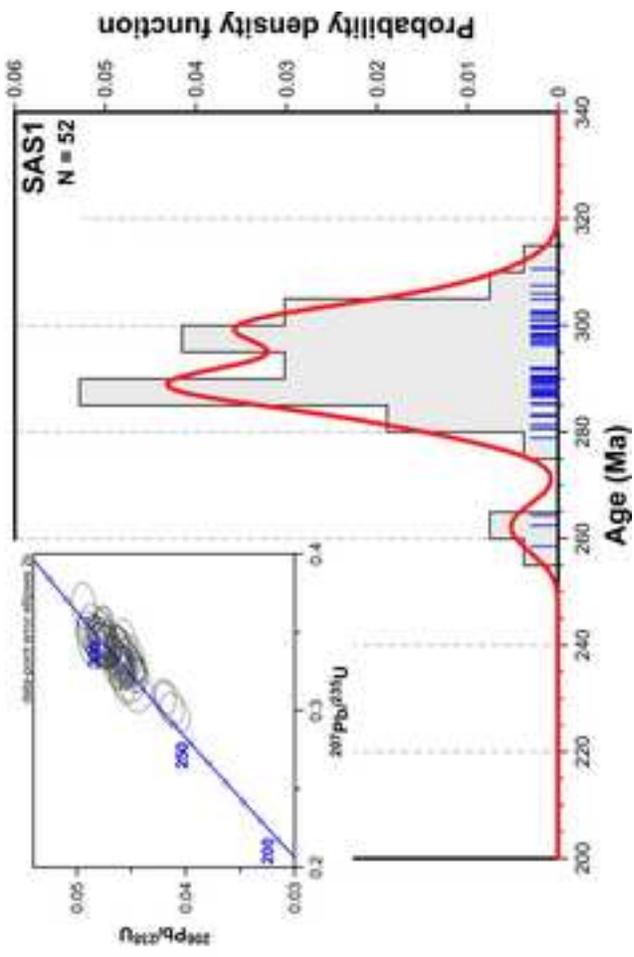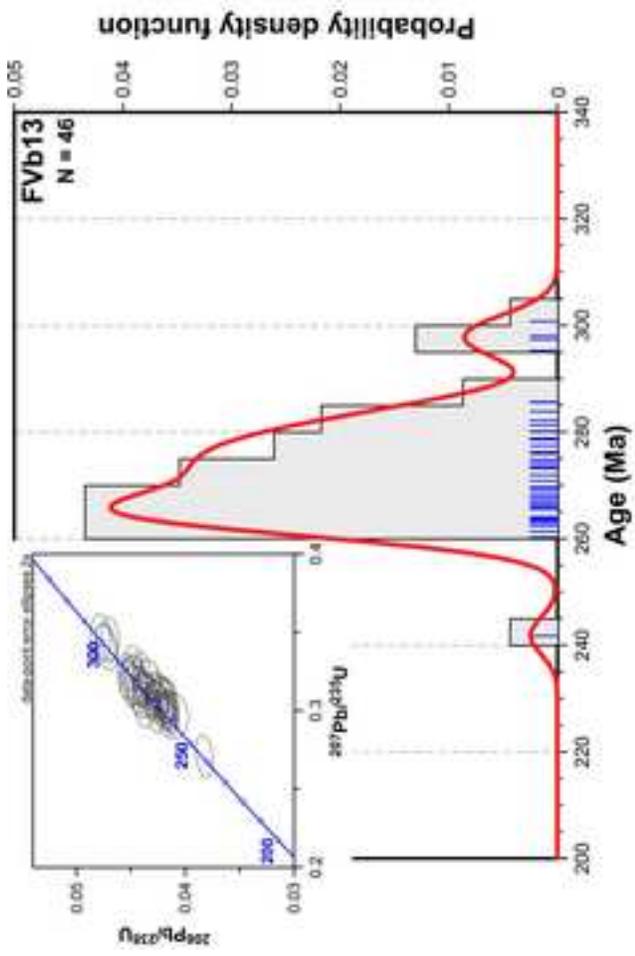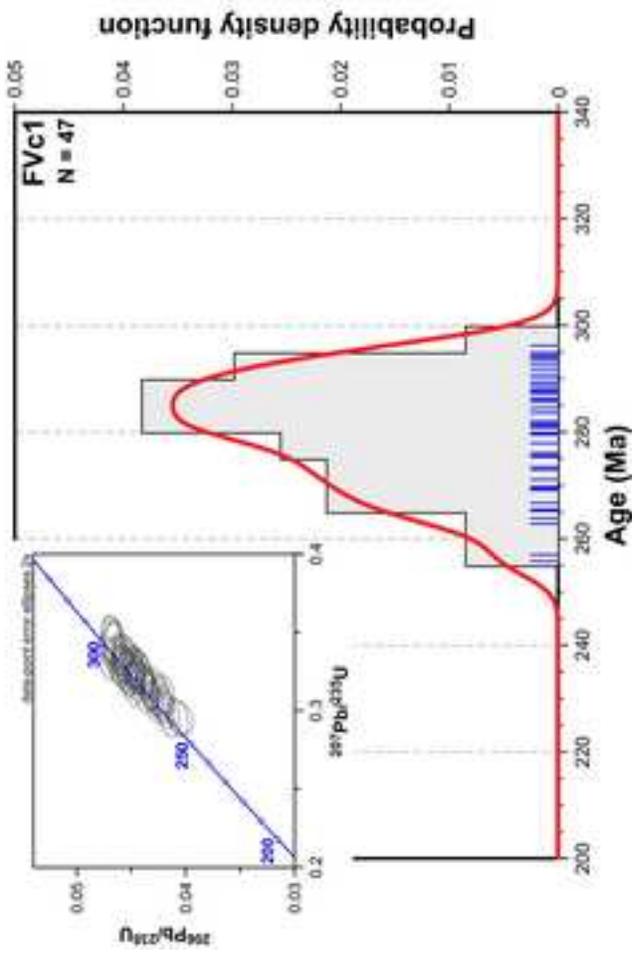

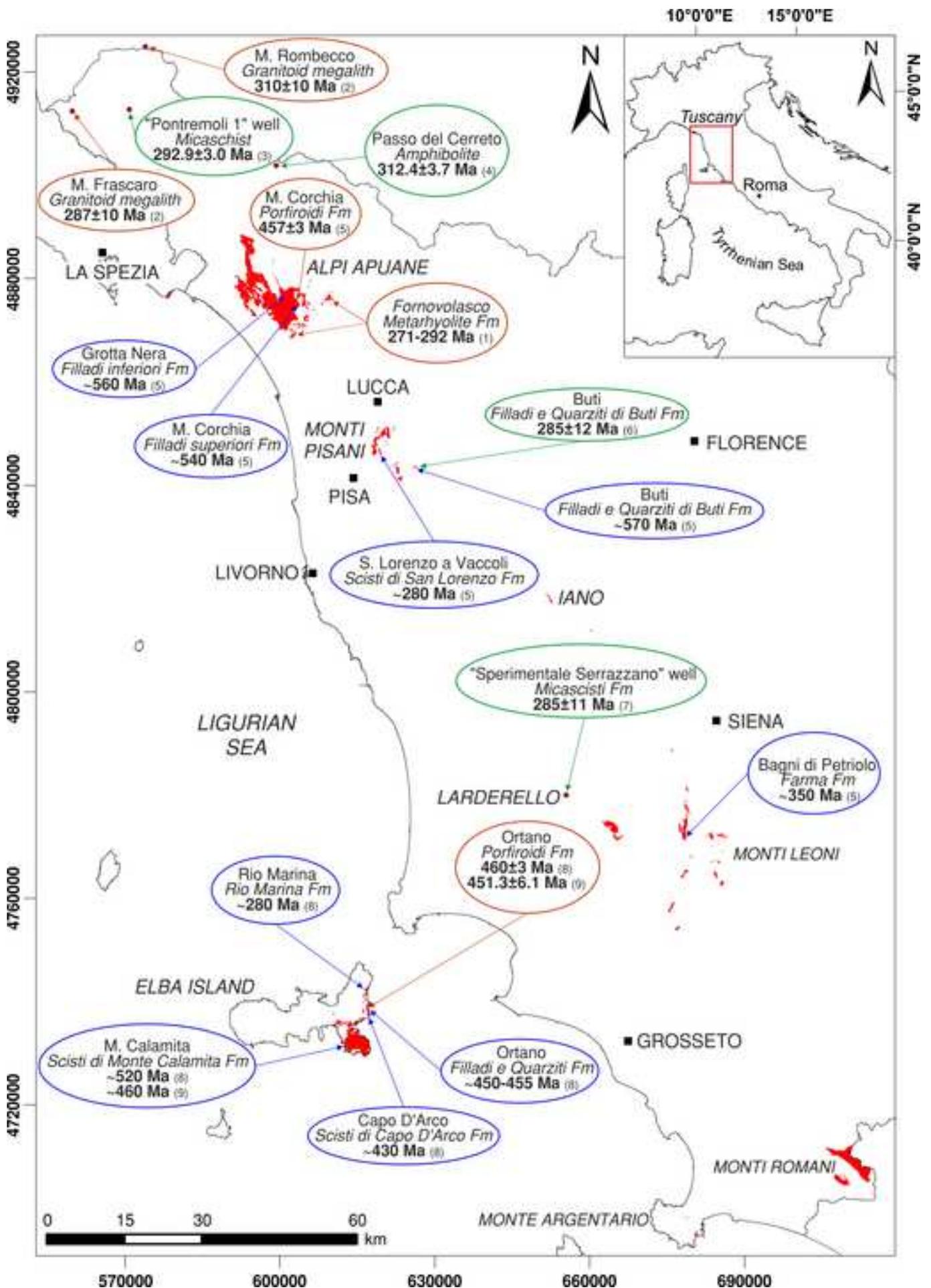